\documentclass[twocolumn,pre,superscriptaddress]{revtex4-1}

\usepackage{bbm}
\usepackage{graphicx}
\usepackage{amsmath,amsfonts,amssymb}
\usepackage{multirow}
\usepackage{rotating}
\usepackage{pseudocode}

\begin{document}
\title{Scoring dynamics across professional team sports: tempo, balance and predictability}

\author{Sears Merritt}
\email[]{sears.merritt@colorado.edu}
\affiliation{Department of Computer Science, University of Colorado, Boulder, CO 80309}

\author{Aaron Clauset}
\email[]{aaron.clauset@colorado.edu}
\affiliation{Department of Computer Science, University of Colorado, Boulder, CO 80309}
\affiliation{BioFrontiers Institute, University of Colorado, Boulder, CO 80303}
\affiliation{Santa Fe Institute, 1399 Hyde Park Rd., Santa Fe, NM 87501}

\begin{abstract}
Despite growing interest in quantifying and modeling the scoring dynamics within professional sports games, relative little is known about what patterns or principles, if any, cut across different sports. Using a comprehensive data set of scoring events in nearly a dozen consecutive seasons of college and professional (American) football, professional hockey, and professional basketball, we identify several common patterns in scoring dynamics. Across these sports, scoring tempo---when scoring events occur---closely follows a common Poisson process, with a sport-specific rate. Similarly, scoring balance---how often a team wins an event---follows a common Bernoulli process, with a parameter that effectively varies with the size of the lead. Combining these processes within a generative model of gameplay, we find they both reproduce the observed dynamics in all four sports and accurately predict game outcomes. These results demonstrate common dynamical patterns underlying within-game scoring dynamics across professional team sports, and suggest specific mechanisms for driving them. We close with a brief discussion of the implications of our results for several popular hypotheses about sports dynamics.
\end{abstract}

\maketitle

Professional team sports like American football, soccer, hockey, basketball, etc.\ provide a rich and relatively well-controlled domain by which to study fundamental questions about the dynamics of competition. In these sports, most environmental irregularities are eliminated, players are highly trained, and rules are enforced consistently. These features produce a level playing field on which competition outcomes are determined largely by a combination of skill and luck (ideally more the former than the latter).

Modern sports in particular produce large quantities of detailed data describing not only competition outcomes and team characteristics, but also the individual events within a competition, e.g., scoring events, referee calls, timeouts, ball possessions, court positions, etc. The availability of such data has enabled many quantitative analyses of individual sports~\cite{klaassen:magnus:2001,albert2005anthology,ben2005most,thomas2007inter,duch:etal:2010,heuer:etal:2010,buttrey2011estimating,radicchi:2011,radicchi:2012,gabel:redner:2012,goldman2012effort,yaari:david:2012}. Relatively little work, however, has asked what patterns or principles, if any, cut across different sports, or whether there are fundamental processes governing some dynamical aspects of all such competitions. These questions are the focus of this study, and our results shed light on several other phenomena, including the roles of skill and luck in determining outcomes, and the extent to which events early in the game influence events later in the game.

Game theory provides an attractive quantitative framework for understanding the principles and dynamics of competition~\cite{myerson:1997}. Given a set of payoffs for different actions, formal game theory can identify the optimal strategy or probability distribution over actions against an intelligent adversary. In simple decision spaces, like penalty shots in soccer~\cite{palacios2003professionals} or serve-and-return play in tennis~\cite{walker2001minimax}, professional athletes appear to behave as game theory predicts (although some do not~\cite{romer:2006}). However, most professional team sports exhibit large and complex decision spaces, with many possible actions of uncertain payoffs, and execution is carried out by an imperfectly coordinated team. Game theory provides less guidance within such complex games, and the resulting dynamics are often better described using tools from dynamical systems~\cite{reed:hughes:2006,galla:farmer:2013}.

Using such an approach, we investigate the within-game scoring dynamics of four team sports, college and professional (American) football, professional hockey, and professional basketball. Our primary goals are (i) to quantify and identify the common empirical patterns in scoring dynamics of these sports, and (ii) to understand the competitive processes that produce these patterns.
We do not consider non-stationary effects across games, e.g., evolving team rosters or skill sets, playing field variables, etc. Instead, we focus explicitly on the sequence of scoring events within games. For each sport, we study three measurable quantities: scoring event tempo, balance, and predictability. We take an inferential approach to investigating their cross-sport patterns and present a generative model of competition dynamics that can be fitted directly to scoring event data within games. We apply this model to a comprehensive data set of 1,279,901 scoring events across 9 or 10 years of consecutive seasons in our four team sports.

There are many claims in both the academic literature and the popular press about scoring dynamics within sports, and sports are often used as exemplars of decision making and dynamics in complex competitive environments~\cite{ayton2004hot,balkundi:harrison:2006,romer:2006,berger:2011}. 
Our results on common patterns in scoring dynamics and the processes that generate them serve to clarify, and in several cases directly contradict, many of these claims, and provide a systematic perspective on the general phenomenon.\\

\paragraph*{Summary of results.}
Across all sports, scoring tempo---when scoring events occur---is remarkably well-described by a Poisson process, in which scoring events occur independently with a sport-specific rate at each second on the game clock. This rate is fairly stable across the course of gameplay, except in the first and last few seconds of a scoring period, where it is much lower or much higher, respectively, than normal.
This common pattern implies that scoring events are largely memoryless, i.e., the timing of events earlier in the game have little or no impact on the timing of future events. Memorylessness contrasts with the dynamics of strategic games like chess or Go, in which events early in a game constrain and drive later events. Instead, professional sports appear to exhibit little strategic entailment, and events are driven instead by short-term optimization for scoring as quickly as possible.

The scoring balance between teams---how often a team wins a scoring event---is well-described by a common Bernoulli process, with a bias parameter that varies effectively over gameplay and across sports. Football and hockey exhibit a common pattern in which the probability of scoring again while in the lead effectively increases with lead size. In basketball, however, this probability decreases with lead size (a phenomenon first identified by~\cite{gabel:redner:2012}). The former pattern is consistent with the outcome of each scoring event being determined by a memoryless coin flip whose bias depends on the difference in the teams' inherent skill levels.
The pattern in basketball is also consistent with such a process, but where on-court team skill varies inversely with lead size as a result of teams deploying their weaker players when they are in the lead and their stronger players when they are not. This player management strategy produces substantially more unpredictable games than in other sports, with winning teams losing their lead and losing teams regaining it much more often than we would normally expect.

Overall, these results reinforce the conclusions from scoring tempo, indicating that event outcomes early in a game have little or no impact on event outcomes later in the game, which reinforces statistical claims that teams do not become ``hot,''~\cite{vergin2000winning,ayton2004hot,gabel:redner:2012} with successes running in streaks. Instead, gameplay is largely a sequence of roughly independent, short-term optimizations aimed at maximizing near-term scoring rates, with little multi-play strategic efforts and few downstream consequences for mistakes or miscalculations. This memorylessness may be caused by a persistently level playing field, which lacks strategically exploitable environmental features~\cite{merritt2013environmental} and forbids actions that might produce sustained competitive advantages~\cite{barney:1991} as a result of within-game choices, e.g., eliminating an opposing team's best players. Table~\ref{table:q:and:a} summarizes these results as they relate to a series of specific questions about scoring dynamics.

We combine these insights within a generative model of gameplay and demonstrate that it accurately reproduces the observed evolution of lead-sizes over the course of games in all four sports, and also makes highly accurate predictions of game outcomes, when only the first few scoring events have occurred. Cursory comparisons suggest that this model achieves accuracy comparable to or better than several commercial odds-makers, despite this model knowing nothing about teams, players, or strategies, and instead relying exclusively on the observed tempo and balance patterns in scoring events.

\section{A null model for competition dynamics}
We first introduce the limiting case of an \textit{ideal competition}, which provides a useful tool by which to identify and quantify interesting deviations within real data, and to generate hypotheses as to what underlying processes might produce them. Although we describe this model in terms of two teams accumulating points, it can in principle be generalized to other forms of competition.

In an ideal competition, events unfold on a perfectly neutral or ``level'' playing field, in which there are no environmental features that could give one side a competitive advantage over the other~\cite{merritt2013environmental}. Furthermore, each side is perfectly skilled, i.e., they possess complete information both about the state of the game, e.g., the position of the ball, the location of the players, etc.\ and the set of possible strategies, their optimum responses, and their likelihood of being employed. This is an unrealistic assumption, as real competitors are imperfectly skilled, and possess both imperfect information and incomplete strategic knowledge of the game. However, increased skill generally implies improved performance on these characteristics, and the limiting case would be perfect skill. Finally, each side exhibits a slightly imperfect ability to execute any particular chosen strategy, which captures the fact that no side can control all variables on the field. In other words, two perfectly skilled teams competing on a level playing field will produce scoring events by chance alone, e.g., a slight miscalculation of velocity, a fumbled pass, shifting environmental variables like wind or heat, etc.

\begin{table*}[t!]
\begin{center}
\begin{tabular}{ll|cc|cc}
sport & abbrv. & seasons & teams & competitions & scoring events \\ \hline
Football (college) & CFB & 10, 2000--2009 & 486 & 14,588 & \hspace{0.7em}120,827 \\
Football (pro) & NFL & 10, 2000--2009 & \hspace{0.5em}31 & \hspace{0.5em}2,654 & \hspace{1.2em}19,476 \\
Hockey (pro) & NHL & 10, 2000--2009 & \hspace{0.5em}29 & 11,813 &  \hspace{1.2em}44,989 \\
Basketball (pro) & NBA & \hspace{0.5em}9, 2002--2010 & \hspace{0.5em}31 & 11,744 & 1,080,285 \\
\end{tabular}
\end{center}
\caption{Summary of data for each sport, including total number of seasons, teams, competitions, and scoring events.}
\label{tab:data}
\end{table*}

An ideal competition thus eliminates all of the environmental, player, and strategic heterogeneities that normally distinguish and limit a team. The result, particularly from the spectator's point of view, is a competition whose dynamics are fundamentally unpredictable. Such a competition would be equivalent to a simple stochastic process, in which scoring events arrive randomly, via a Poisson process with rate $\lambda$, points are awarded to each team with equal probability, as in a fair Bernoulli process with parameter $c=1/2$, and the number of those points is an iid random variable from some sport-specific distribution.

Mathematically, let $S_r(t)$ and $S_b(t)$ denote the cumulative scores of teams $r$ and $s$ at time $t$, where $ 0 \le t \le T$ represents the game clock. (For simplicity, we do not treat overtime and instead let the game end at $t=T$.) The probability that $S_r$ increases by $k$ points at time $t$ is equal to the joint probability of observing an event worth $k$ points, scored by team $r$ at time $t$. Assuming independence, this probability is 
\begin{equation}
\Pr(\Delta S_r(t) = k) = \Pr(\textrm{event at $t$})\Pr(\textrm{$r$ scores})\Pr(\textrm{points $=k$}) \enspace .
\end{equation}
The evolution of the difference in these scores thus follows an finite-length unbiased random walk on the integers, moving left or right with equal probability, starting at $\Delta S=0$ at $t=0$.

Real competitions will deviate from this ideal because they possess various non-ideal features. The type and size of such deviations are evidence for competitive mechanisms that drive the scoring dynamics away from the ideal.

\section{Scoring event data}
Throughout our analyses, we utilize a comprehensive data set of all points scored in league games of consecutive seasons of college-level American football (NCAA Divisions 1--3, 10 seasons; 2000--2009), professional American football (NFL, 10 seasons; 2000--2009), professional hockey (NHL, 10 seasons; 2000--2009), and professional basketball (NBA, 9 seasons, 2002--2010).%
\footnote{Data provided by STATS LLC, copyright 2014.}
Each scoring event includes the time at which the event occurred, the player and corresponding team that won the event, and the number of points it was worth. From these, we extract all scoring events that occurred during regulation time (i.e., we exclude all overtime events), which account for 99\% or more of scoring events in each sport, and we combine events that occur at the same second of game time. Table~\ref{tab:data} summarizes these data, which encompass more than 1.25 million scoring events across more than 40,000 games.

A brief overview of each sport's primary game mechanics is provided in Additional File~1 as Appendix~A. In general, games in these sports are competitions between two teams of fixed size, and points are accumulated each time one team places the ball or puck in the opposing team's goal. Playing fields are flat, featureless surfaces. Gameplay is divided into three or four scoring periods within a maximum of 48 or 60 minutes (not including potential overtime). The team with the greatest score at the end of this time is declared the winner.

\section{Game tempo}
A game's ``tempo'' is the speed at which scoring events occur over the course of play. Past work on the timing of scoring events has largely focused on hockey, soccer and basketball~\cite{thomas2007inter,heuer:etal:2010,gabel:redner:2012}, with little work examining football or in contrasting patterns across sports. However, these studies show strong evidence that game tempo is well approximated by a homogenous Poisson process, in which scoring events occur at each moment in time independently with some small and roughly constant probability. 

Analyzing the timing of scoring events across all four of our sports, we find that the Poisson process is a remarkably good model of game tempo, yielding predictions that are in good or excellent agreement with a variety of statistical measures of game play. Furthermore, these results confirm and extend previous work~\cite{ayton2004hot,gabel:redner:2012}, while contrasting with others~\cite{yaari:david:2012,yaari:eisenman:2011}, showing little or no evidence for the popular belief in ``momentum'' or ``hot hands,'' in which scoring once increases the probability of scoring again very soon. However, we do find some evidence for modest non-Poissonian patterns in tempo, some of which are common to all four sports. 

\begin{table}[b!]
\begin{center}
\begin{tabular}{l|cc|cc}
sport & $\hat{\lambda}$  & $T$ & $\hat{\lambda} T$ & $1/\hat{\lambda}$  \\ 
& [events / s] & [s] & [events / game] & [s / event] \\ \hline
NFL & 0.00204(1) & 3600 & \hspace{0.5em}7.34 & 490.2 \\
CFB & 0.00230(1) & 3600 & \hspace{0.5em}8.28 & 434.8 \\
NHL & 0.00106(1) & 3600 & \hspace{0.5em}3.81 & 943.4 \\
NBA & 0.03194(5) & 2880 & 91.99 & \hspace{0.5em}31.3 
\end{tabular}
\end{center}
\caption{Tempo summary statistics for each sport, along with simple derived values for the expected number of events per game and seconds between events. Parenthetical values indicate standard uncertainty in the final digit.}
\label{table:tempo:balance}
\end{table}

\begin{figure*}[t!]
\centering
\includegraphics[scale=0.99]{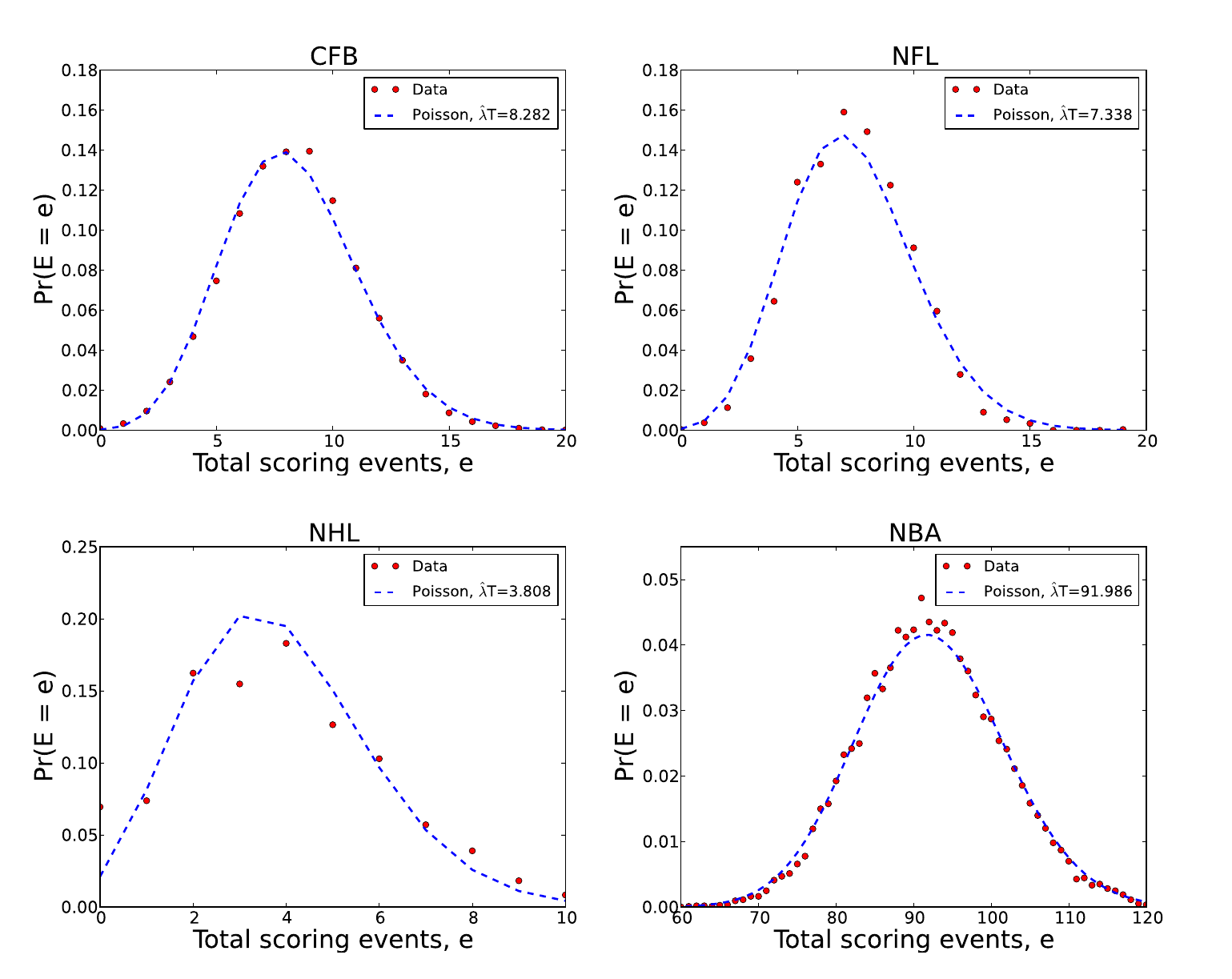}
\caption{\textbf{Scoring events per game.} Empirical distributions for the number of scoring events per game, along with the estimated Poisson model with rate $\lambda T$ (dashed).}
\label{fig:timing:events}
\end{figure*}

\subsection{The Poisson model of tempo}
A Poisson process is fully characterized by a single parameter $\lambda$, representing the probability that an event occurs, or the expected number of events, per unit time. In each sport, game time is divided into seconds and there are $T$ seconds per a game (see Table~\ref{table:tempo:balance}). For each sport, we test this model in several ways: we compare the empirical and predicted distributions for the number of events per game and for the time between consecutive scoring events, and we examine the two-point correlation function for these inter-event times.

Under a Poisson model~\cite{boas:2006}, the number of scoring events per game follows a Poisson distribution with parameter $\lambda T$, and the maximum likelihood estimate of $\lambda$ is the average number of events observed in a game divided by the number of intervals (which varies per sport). Furthermore, the time between consecutive events follows a simple geometric (discrete exponential) distribution, with mean $1/\lambda$, and the two-point correlation between these delays is zero at all time scales.

For the number of events per game, we find generally excellent agreement between the Poisson model and the data for every sport (Figure~\ref{fig:timing:events}). However, there are some small deviations, which suggests some second-order, non-Poissonian processes, which we investigate below. Deviations are greatest in NHL games, whose distribution is slightly broader than predicted, underproducing games with 3 events, and overproducing games with 0 or with 8 or more events. Similarly, CFB games have a slight excess of games with 9 events, and NBA games exhibit slightly more variation in NBA games with scores close to the average (92.0 events) than expected. In contrast, NFL games exhibit slightly less variance than expected, with more games close to the average (7.3 events) than expected.

\begin{figure*}[t!]
\centering
\includegraphics[scale=0.99]{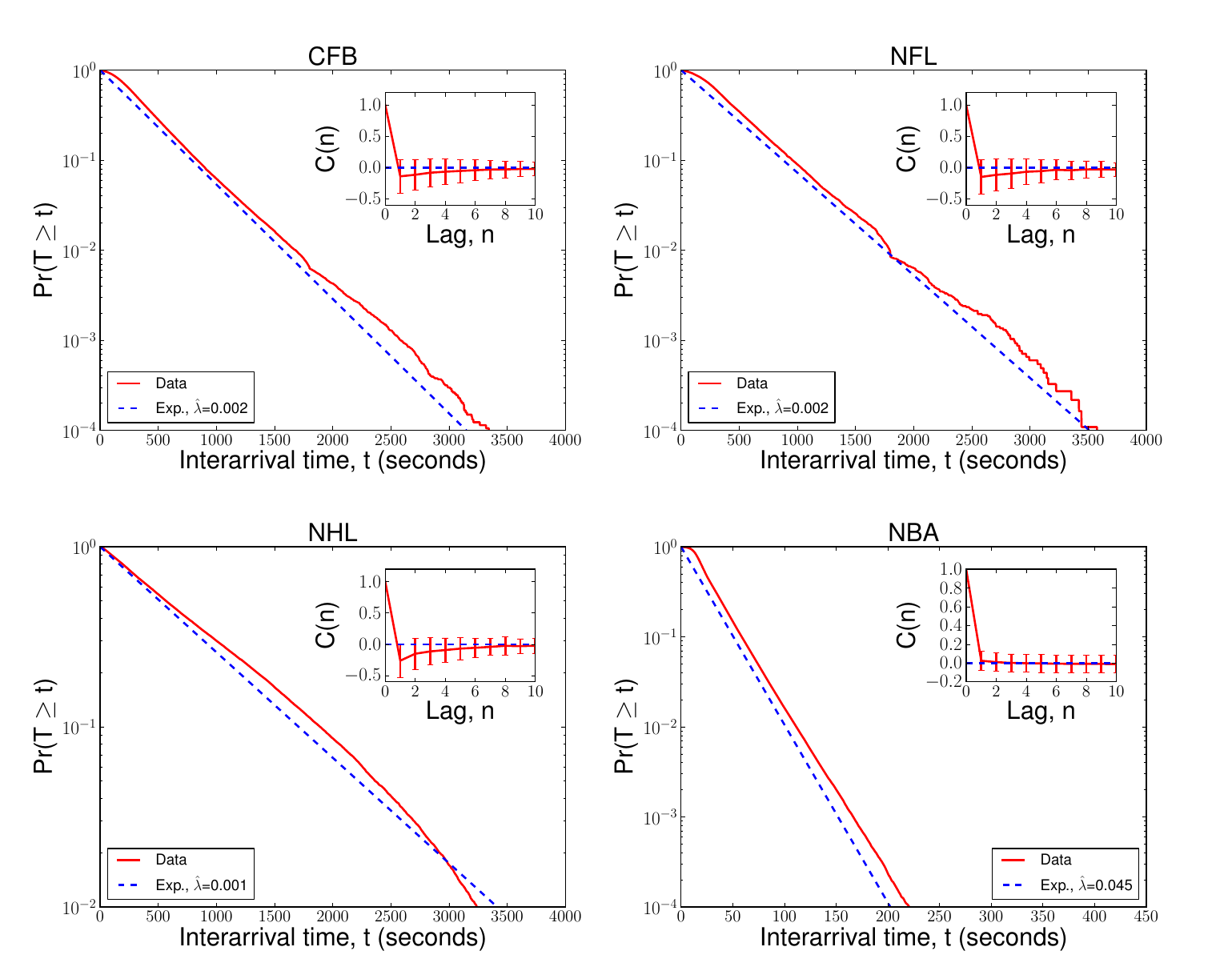}
\caption{\textbf{Time between scoring events.} Empirical distribution of time between consecutive scoring events, shown as the complementary cdf, along with the estimated distribution from the Poisson model (dashed). Insets show the correlation function for inter-event times.}
\label{fig:timing:iat}
\end{figure*}

For the time between consecutive scoring events within a game, or the inter-arrival time distribution, we again find excellent agreement between the Poisson model and the data in all sports (Figure~\ref{fig:timing:iat}). That being said, in CFB, NFL and NBA games, there are slightly fewer gaps of the minimum size than predicted by the model. This indicates a slight dispersive effect in the timing of events, perhaps caused by the time required to transport the ball some distance before a new event may be generated. In contrast, NHL games produce as many short gaps, more intermediate gaps, and fewer very long gaps than expected were events purely Poissonian.

Finally, we calculate the two-point correlation function on the times between scoring events~\cite{box2011time},
\begin{equation}
C(n) = \left( \sum_k (t_k - \langle t \rangle)(t_{k+n}-\langle t \rangle) \right) \left/  \sum_k (t_k - \langle t\rangle)^2  \right. \enspace ,
\end{equation}
where $t_{k}$ is the $k$th inter-arrival time, $n$ indicates the gap between it and a subsequent event, and $\langle t \rangle$ is the mean time between events. If $C(n)$ is positive, short intervals tend to be followed by other short intervals (or, large intervals by large intervals), while a negative value implies alternation, with short intervals followed by long, or vice versa. Across all four sports, the correlation function is close or very close to zero for all values of $n$ (Figure~\ref{fig:timing:iat} insets), in excellent agreement with the Poisson process, which predicts $C(n)=0$ for all $n>0$, representing no correlation in the timing of events (a result also found by~\cite{gabel:redner:2012} in basketball). However, in CFB, NFL and NHL games, we find a slight negative correlation for very small values of $n$, suggesting a slight tendency for short intervals to be closely followed by longer ones, and vice versa.

\begin{figure*}[t!]
\centering
\includegraphics[scale=0.99]{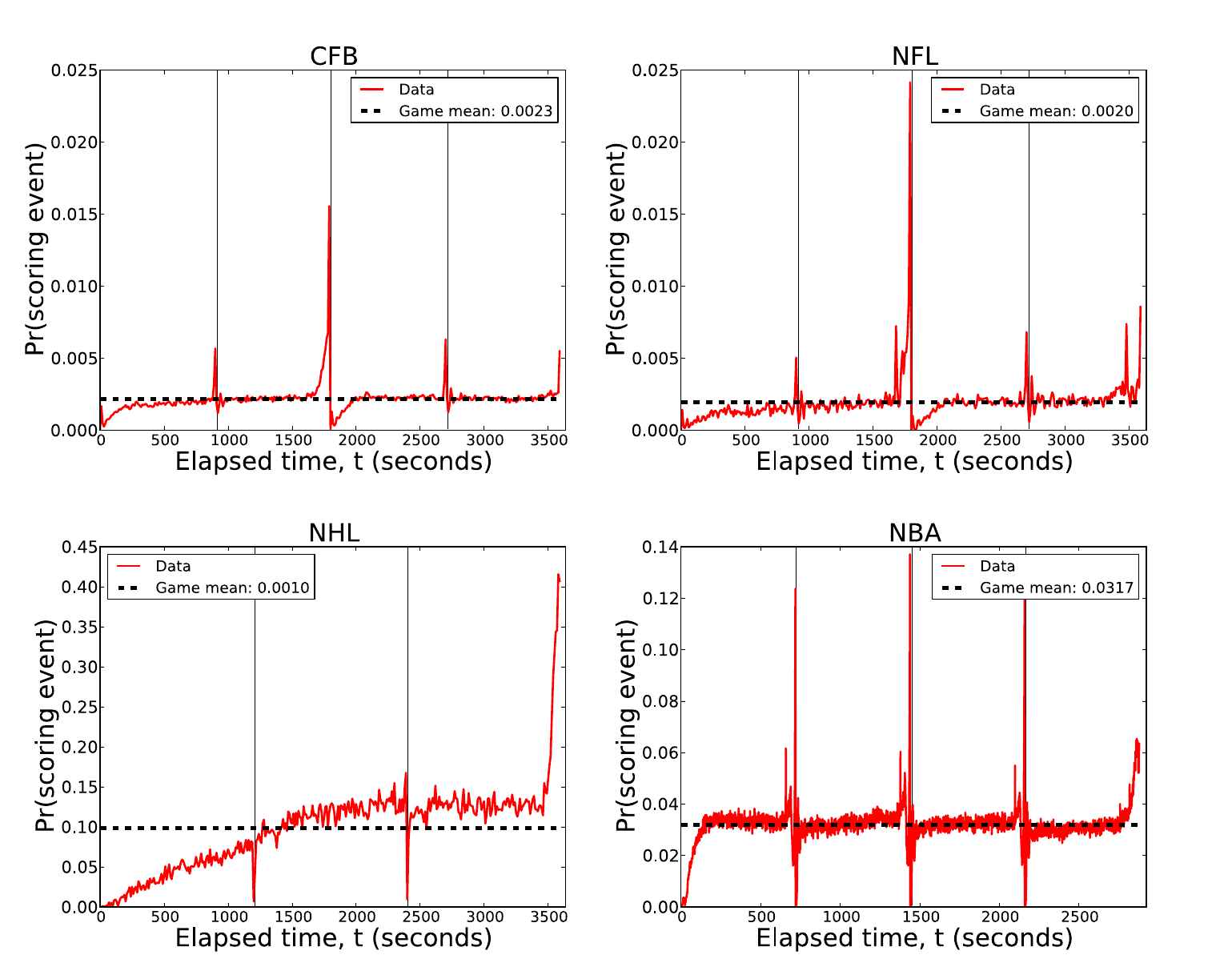}
\caption{\textbf{Game tempo.} Empirical probability of scoring events as a function of game time, for each sport, along with the mean within-sport probability (dashed line). Each distinct game period, demarcated by vertical lines, shows a common three-phase pattern in tempo.}
\label{fig:timing:rate}
\end{figure*}

\subsection{Common patterns in game tempo}
\label{sec:tempo}
Our results above provide strong support for a common Poisson-like process for modeling game tempo across all four sports. We also find some evidence for mild non-Poissonian processes, which we now investigate by directly examining the scoring rate as a function of clock time. Within each sport, we tabulate the fraction of games in which a scoring event (associated with any number of points) occurred in the $t$th second of gameplay.

Across all sports, we find that the tempo of events follows a common three-phase pattern within each distinct period of play (Figure~\ref{fig:timing:rate}). This pattern, which resembles an inverse sigmoid, is characterized by (i) an early phase of non-linearly increasing tempo, (ii) a middle phase of stable (Poissonian) or slightly increasing tempo, and (iii) an end phase of sharply increasing tempo. This pattern is also observed in certain online games~\cite{merritt2013environmental}, which have substantially different rules and are played in highly heterogeneous environments, suggesting a possibly fundamental generating mechanism for team-competitive systems. 

\paragraph*{Early phase:\ non-linear increase in tempo.} When a period begins, players are in specific and fixed locations on the field, and the ball or puck is far from any team's goal. Thus, without regard to other aspects of the game, it must take some time for players to move out of these initial positions and to establish scoring opportunities. This would reduce the probability of scoring relative to the game average by limiting access to certain player-ball configurations that require time to set up. Furthermore, and potentially most strongly in the first of these phases (beginning at $t=0$), players and teams may still be ``warming up,'' in the sense of learning~\cite{thompson:2010} the capabilities and tendencies of the opposing team and players, and which tactics to deploy against the opposing team's choices. These behaviors would also reduce the probability of scoring by encouraging risk averse behavior in establishing and taking scoring opportunities.

We find evidence for both mechanisms in our data. Both CFB and NFL games exhibit short and modest-sized dips in scoring rates in periods 2 and 4, reflecting the fact that player and ball positions are not reset when the preceding quarters end, but rather gameplay in the new quarter resumes from its previous configuration. In contrast, CFB and NFL periods 1 and 3 show significant drops in scoring rates, and both of these quarters begin with a kickoff from fixed positions on the field. Similarly, NBA and NHL games exhibit strong but short-duration dips in scoring rate at the beginning of each of their periods, reflecting the fact that each quarter begins with a tossup or face-off, in which players are located in fixed positions on the court or rink. NBA and football games also exhibit some evidence of the ``warming up'' process, with the overall scoring rate being slightly lower in period 1 than in other equivalent periods. In contrast, NHL games exhibit a prolonged warmup period, lasting well past the end of the first period. This pattern may indicate more gradual within-game learning in hockey, perhaps are a result of the large diversity of on-ice player combinations caused by teams rotating their four ``lines'' of players every few minutes.

\paragraph*{Middle phase:\ constant tempo.} Once players have moved away from their initial locations and/or warmed up, gameplay proceeds fluidly, with scoring events occurring without any systematic dependence on the game clock. This produces a flat, stable or stationary pattern in the probability of scoring events. A slight but steady increase in tempo over the course of this phase is consistent with learning, perhaps as continued play sheds more light on the opposing team's capabilities and weaknesses, causing a progressive increase in scoring rate as that knowledge is accumulated and put into practice.

A stable scoring rate pattern appears in every period in NFL, CFB and NBA games, with slight increases observed in periods 1 and 2 in football, and in periods 2--4 in basketball. NHL games exhibit stable scoring rates in the second half of period 2 and throughout period 3. Within a given game, but across scoring periods, scoring rates are remarkably similar, suggesting little or no variation in overall strategies across the periods of gameplay.

\paragraph*{End phase:\ sharply increased tempo.} The end of a scoring period often requires players to reset their positions, and any effort spent establishing an advantageous player configuration is lost unless that play produces a scoring event. This impending loss-of-position will tend to encourage more risky actions, which serve to dramatically increase the scoring rate just before the period ends. The increase in scoring rate should be largest in the final period, when no additional scoring opportunities lay in the future. In some sports, teams may effectively slow the rate by which time progresses through game clock management (e.g., using timeouts) or through continuing play (at the end of quarters in football). This effectively compresses more actions than normal into a short period of time, which may also increase the rate, without necessarily adding more risk.

We find evidence mainly for the loss-of-position mechanism, but the rules of these games suggest that clock management likely also plays a role. Relative to the mean tempo, we find a sharply increased rate at the end of each sport's games, in agreement with a strong incentive to score before a period ends. (This increase indicates that a ``lolly-gag strategy,'' in which a leading team in possession intentionally runs down the clock to prevent the trailing team from gaining possession, is a relatively rare occurrence.) Intermediate periods in NFL, CFB and NBA games also exhibit increased scoring rates in their final seconds. In football, this increase is greatest at the end of period 2, rather than period 4. The increased rate at the ends of periods 1 and 3 in football is also interesting, as here the period's end does not reset the player configuration on the field, but rather teams switch goals. This likely creates a mild incentive to initiate some play before the period ends (which is allowed to finish, even if the game clock runs out). NHL games exhibit no discernible end-phase pattern in their intermediate periods (1 and 2), but show an enormous end-game effect, with the scoring rate growing to more than three times its game mean. This strong pattern may be related to the strategy in hockey of the losing team ``pulling the goalie,'' in which the goalie leaves their defensive position in order to increase the chances of scoring. Regardless of the particular mechanism, the end-phase pattern is ubiquitous.

In general, we find a common set of modest non-Poissonian deviations in game tempo across all four sports, although the vast majority of tempo dynamics continue to agree with a simple Poisson model.

\section{Game balance}
A game's ``balance'' is the relative distribution of scoring events (not points) among the teams. Perfectly balanced games, however, do not always result in a tie. In our model of competition, each scoring event is awarded to one team or the other by a Bernoulli process, and in the case of perfect balance, the probability is equal, at $c=1/2$. The expected fraction of scoring events won by a team is also $c=1/2$, and its distribution depends on the number of scoring events in the game. We estimate this null distribution by simulating perfectly balanced games for each sport, given the empirical distribution of scoring events per game (see Fig.~\ref{fig:timing:events}). Comparing the simulated distribution against the empirical distribution of $c$ provides a measure of the true imbalance among teams, while controlling for the stochastic effects of events within games. 

Across all four sports, we find significant deviations in this fraction relative to perfect balance. NFL and CFB games exhibited more variance than expected, while NHL and NBA games exhibited the least. Within a game, scoring balance exhibits unexpected patterns. In particular NBA games exhibit an unusual ``restoring force'' pattern, in which the probability of winning the next scoring event \textit{decreases} with the size of a team's lead (a pattern first observed by~\cite{gabel:redner:2012}). In contrast, NFL, CFB and NHL games exhibit the opposite effect, in which the probability of winning the next scoring event appears to increase with the size of the lead---a pattern consistent with a heterogeneous distribution of team skill.

\begin{figure*}[t]
\centering
\includegraphics[scale=0.99]{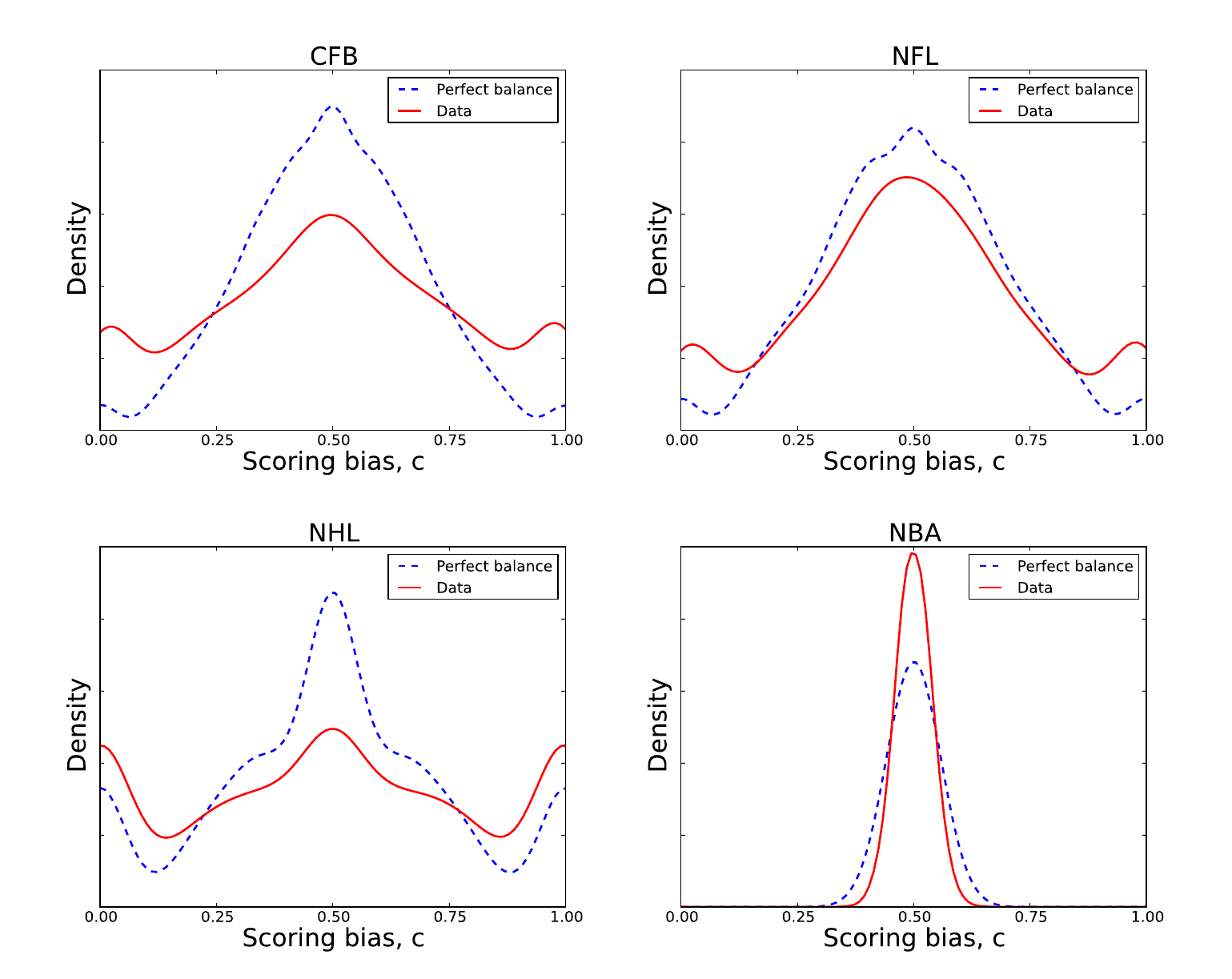}
\caption{\textbf{Game balance.} Smoothed distributions for the empirical fraction $\hat{c}$ of events won by a team, for each sport, and the predicted fraction for a perfectly balanced scoring, when given the empirical distribution of events per game (Fig.~\ref{fig:timing:events}). Modes at 1 and 0 indicate a non-trivial probability of one team winning or losing every event, which is more common when only a few events occur.}
\label{fig:scoring:balance}
\end{figure*}

\subsection{Quantifying balance}
The fraction of all events in the game that were won by a randomly selected team provides a simple 
measure of the overall balance of a particular game in a sport. Let $r$ and $b$ index the two teams and let $E_{r}$ ($E_{b}$) denote the total number of events won by team $r$ in its game with $b$. The maximum likelihood estimator for a game's bias is simply the fraction $\hat{c} = E_r\left/ (E_r+E_b) \right.$ of all scoring events in the game won by $r$.

Tabulating the empirical distributions of $\hat{c}$ within each sport, we find that the most common outcome, in all sports, is $c=1/2$, in agreement with the Bernoulli model. However, the distributions around this value deviate substantially from the form expected for perfect balance (Figure~\ref{fig:scoring:balance}), but not always in the same direction.

In CFB and NFL, the distributions of scoring balances are similar, but the shape for CFB is broader than for NFL, suggesting that CFB competitions are less balanced than NFL competitions. This is likely a result of the broader range of skill differences among teams at the college level, as compared to the professionals. Like CFB and NFL, NHL games also exhibit substantially more blowouts and fewer ties than expected, which is consistent with a heterogeneous distribution of team skills. Surprisingly, however, NBA games exhibit less variance in the final relative lead size than we expect for perfectly balanced games, a pattern we will revisit in the following section.

\subsection{Scoring while in the lead}
Although many non-Bernoulli processes may occur within professional team sports, here we examine only one: whether the size of a lead $L$, the difference in team scores or point totals, provides information about the probability of a team winning the next event. \cite{gabel:redner:2012} previously considered this question for scoring events and lead sizes within NBA games, but not other sports. Across all four of our sports, we tabulated the fraction of times the leading team won the next scoring event, given it held a lead of size $L$. This function is symmetric about $L=0$, where it passes through probability $p=1/2$ where the identity of the leading team may change.

\begin{figure*}[t]
\centering
\includegraphics[scale=0.99]{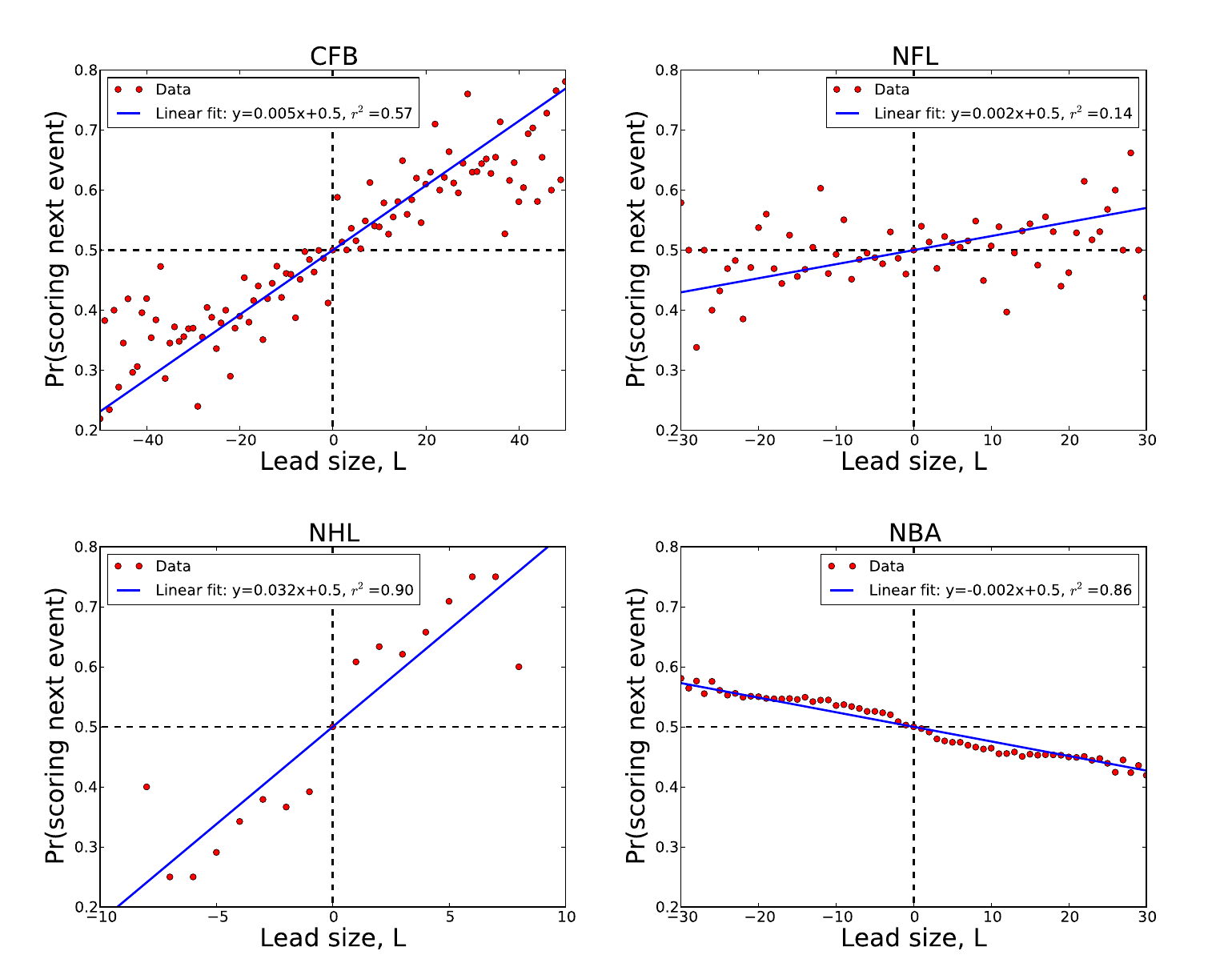}
\caption{\textbf{Lead-size dynamics.} The probability of scoring as a function of a team's lead size for each sport, football, hockey, and basketball and a linear least-squares fit ($p\le0.1$), indicating positive or negative correlations between scoring and a competition's score difference.}
\label{fig:scoring:lead}
\end{figure*}

Examining the empirical scoring functions (Figure~\ref{fig:scoring:lead}), we find that the probability of scoring next varies systematically with lead size $L$. In particular, for CFB, NFL and NHL games, the probability appears to increase with lead size, while it decreases in NBA games. The effect of the negative relationship in NBA games is a kind of ``restoring force,'' such that leads of any size tend to shrink back toward a tied score. This produces a narrower distribution of final lead sizes than we would expect under Bernoulli-style competition, precisely as shown in Figure~\ref{fig:scoring:balance} for NBA games.

Although the positive function for CFB, NFL and NHL games may superficially support a kind of ``hot hands'' or cumulative advantage-type mechanism, in which lead size tends to grow superlinearly over time, we do not believe this explains the observed pattern. A more plausible mechanism is a simple heterogeneous skill model, in which each team has a latent skill value $\pi_{r}$, and the probability that team $r$ wins a scoring event against $b$ is determined by a Bernoulli process with $c=\pi_{r}\left/ (\pi_{r}+\pi_{b})\right.$. (This model is identical to the popular Bradley-Terry model of win-loss records of teams~\cite{bradley1952rank}, except here we apply it to each scoring event within a game.)

For a broad class of team-skill distributions, this model produces a scoring function with the same sigmoidal shape seen here, and the linear pattern at $L=0$ is the result of averaging over the distribution of biases $c$ induced by the team skill distribution. The function flattens out at large $|L|$ assuming the value representing the largest skill difference possible among the league teams. This explanation is supported by the stronger correlation in CFB games ($+0.005$ probability per point in the lead) versus NFL games ($+0.002$ probability per point), as CFB teams are known to exhibit much broader skill differences than NFL teams, in agreement with our results above in Figure~\ref{fig:scoring:balance}.

NBA games, however, present a puzzle, because no distribution of skill differences can produce a negative correlation under this latent-skill model. \cite{gabel:redner:2012} suggested this negative pattern could be produced by possession of the ball changing after each scoring event, or by the leading team ``coasting'' and thereby playing below their true skill level. However, the change-of-possession rule also exists in CFB and NFL games (play resumes with a faceoff in NHL games), but only NBA games exhibit the negative correlation. Coasting could occur for psychological reasons, in which losing teams play harder, and leading teams less hard, as suggested by~\cite{berger:2011}. Again, however, the absence of this pattern in other sports suggest that the mechanism is not psychological.

\begin{figure*}[t]
\centering
\includegraphics[scale=0.99]{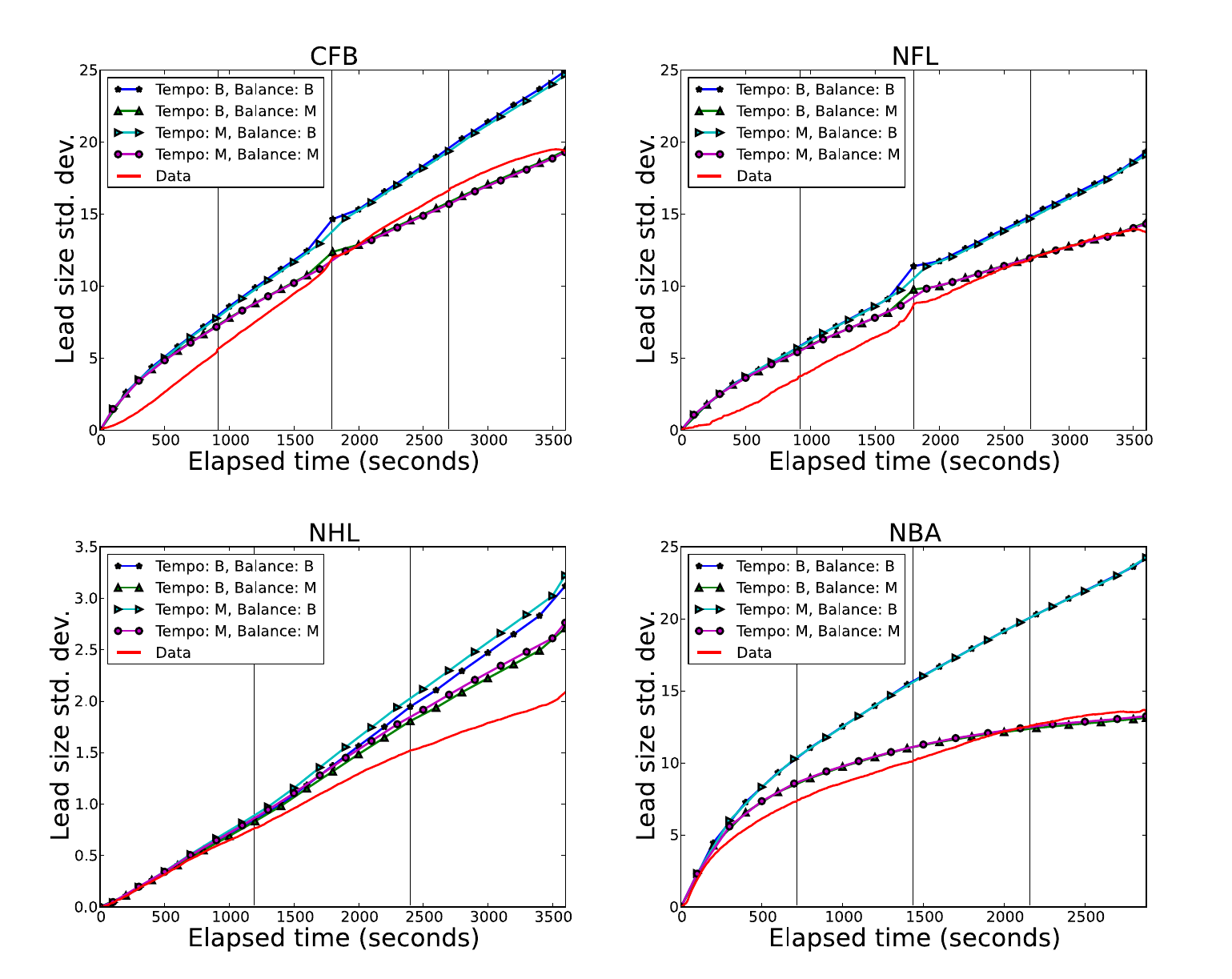}
\caption{\textbf{Modeling lead-size dynamics.} Comparison of empirical lead-size variation as a function of clock time with those produced by  Bernoulli (B) or Markov (M) tempo or balance models, for each sport.}
\label{fig:sim:lead}
\end{figure*}

A plausible alternative explanation is that NBA teams employ various strategies that serve to change the ratio $c=\pi_{r}\left/ (\pi_{r}+\pi_{b})\right.$ as a function of lead size. For instance, when a team is in the lead, they often substitute out their stronger and more offensive players, e.g., to allow them to rest or avoid injury, or to manage floor spacing or skill combinations. When a team is down by an amount that likely varies across teams, these players are put back on the court. If both teams pursue such strategies, then effective ratio $c$ will vary inversely with lead size such that the leading team becomes effectively weaker compared to the non-leading team. In contrast to NBA teams, teams in CFB, NFL and NHL seem less able to pursue such a strategy. In football, substitutions are relatively uncommon, implying that $\pi_{r}$ should not vary much over the course of a game. In hockey, each team rotates through most of its players every few minutes, which limits the ability for high- or low-skilled players to effectively change $\pi_{r}$ over the course of a game.

\section{Modeling lead-size dynamics}
\label{sec:lead:size:dynamics}
The previous insights identify several basic patterns in scoring tempo and balance across sports. However, we still lack a clear understanding of the degree to which any of these patterns is necessary to produce realistic scoring dynamics. Here, we investigate this question by combining the identified patterns within a generative model of scoring over time, and test which combinations produce realistic dynamics in lead sizes. In particular, we consider two models of tempo and two models of balance. For each of the four pairs of tempo and balance models for each sport, we generate via Monte Carlo a large number of games and measure the resulting variation in lead size as a function of the game clock, which we then compare to the empirical pattern.

Our two scoring tempo models are as follows. In the first (Bernoulli) model, each second of time produces an event with the empirical probability observed for that second across all games (shown in Figure~\ref{fig:timing:rate}). In the second (Markov), we draw an inter-arrival time from the empirical distribution of such gaps (shown in Figure~\ref{fig:timing:iat}), advance the game clock of that amount, and generate a scoring event at that clock time.

Our two balance models are as follows. In the first (Bernoulli) model, for each match we draw a uniformly random value $c$ from the empirical distribution of scoring balances (shown in Figure~\ref{fig:scoring:balance}) and for each scoring event, the points are won by team $r$ with that probability and by team $b$ otherwise. In the second (Markov), a scoring event is awarded to the leading team with the empirically estimated probability for the current lead size $L$ (shown in Figure~\ref{fig:scoring:lead}). Once a scoring event is generated and assigned, that team's score is incremented by a point value drawn iid from the empirical distribution of point values per scoring event for the sport (see Appendix~\ref{appendix:points:per:event}).

The four combinations of tempo and balance models thus cover our empirical findings for patterns in the scoring dynamics of these sports. The simpler models (called Bernoulli) represent dynamics with no memory, in which each event is an iid random variable, albeit drawn from a data-driven distribution. The more complicated models (called Markov) represent dynamics with some memory, allowing past events to influence the ongoing gameplay dynamics. In particular, these are first-order Markov models, in which only the events of the most recent past state influence the outcome of the random variable at the current state.

Generating 100,000 competitions under each combination of models for each sport, we find a consistent pattern across sports (Figure~\ref{fig:sim:lead}): the Markov model of game tempo provides little improvement over the Bernoulli model in capturing the empirical pattern of lead-size variation, while the Markov model for balance provides a significant improvement over the Bernoulli model. In particular, the Markov model generates gameplay dynamics in very good agreement with the empirical patterns.

That being said, some small deviations remain. For instance, the Markov model slightly overestimates the lead-size variation in the first half, and slightly underestimates it in the second half of CFB games. In NFL games, it provides a slight overestimate in first half, but then converges on the empirical pattern in the second half. NHL games exhibit the largest and most systematic deviation, with the Markov model producing more variation than observed, particularly in the game's second half. However, it should be noted that the low-scoring nature of NHL means that what appears to be a visually large overestimate here (Fig.~\ref{fig:sim:lead}) is small when compared to the deviations seen in the other sports. NBA games exhibit a similar pattern to CFB games, but the crossover point occurs at the end of period 3, rather than at period 2. These modest deviations suggest the presence of still other non-ideal processes governing the scoring dynamics, particularly in NHL games.

We emphasize that the Markov model's accuracy for CFB, NFL and NHL games does not imply that individual matches follow this pattern of favoring the leader. Instead, the pattern provides a compact and efficient summary of scoring dynamics conditioned on unobserved characteristics like team skill. Our model generates competition between two featureless teams, and the Markov model provides a data-driven mechanism by which some pairs of teams may behave as if they have small or large differences in latent skill. It remains an interesting direction for future work to investigate precisely how player and team characteristics determine team skill, and how team skill impacts scoring dynamics.

\section{Predicting outcomes from gameplay}
The accuracy of our generative model in the previous section suggest that it may also produce accurate predictions of the game's overall outcome, after observing only the events in the first $t$ seconds of the game. In this section, we study the predictability of game outcome using the Markov model for scoring balance, and compare its accuracy to the simple heuristic of guessing the winner to be the team currently in the lead at time $t$. Thus, we convert our Markov model into an explicit Markov chain on the lead size $L$, which allows us to simulate the remaining $T-t$ seconds conditioned on the lead size at time $t$. For concreteness, we define the lead size $L$ relative to team $r$, such that $L<0$ implies that $b$ is in the lead.

The Markov chain's state space is the set of all possible lead sizes (score differences between teams $r$ and $b$), and its transition matrix $P$ gives the probability that a scoring event changes a lead of size $L$ to one of size $L'$. If $r$ wins the event, then $L'=L+k$, where $k$ is the event's point value, while if $b$ wins the event, then $L'=L-k$. Assuming the value and winner of the event are independent, the transition probabilities are given by
\begin{align}
P_{L,L+k} & = \Pr(\textrm{$r$ scores $ |\, L$}) \Pr(\textrm{point value $= k$}) \nonumber \\ 
P_{L,L-k} & = (1-\Pr(\textrm{$r$ scores $ |\, L$})) \Pr(\textrm{point value $= k$}) \enspace , \nonumber 
\end{align}
where, for the particular sport, we use the empirical probability function for scoring as a function of lead size (Figure~\ref{fig:scoring:lead}), from $r$'s perspective, and the empirical distribution (Appendix~\ref{appendix:points:per:event}) for the point value.

The probability that team $r$ is the predicted winner depends on the probability distribution over lead sizes at time $T$. Because scoring events are conditionally independent, this distribution is given by $P^{n}$, where $n$ is the expected number of scoring events in the remaining clock time $T-t$, multiplied by a vector $S_{0}$ representing the initial state $L=0$. Given a choice of time $t$, we estimate $n  = \sum_{w=t}^T \Pr(\textrm{event}\,|\,w)$, which is the expected number of events given the empirical tempo function (Fig.~\ref{fig:timing:rate}, also the Bernoulli tempo model in Section~\ref{sec:lead:size:dynamics}) and the remaining clock time.
We then convert this distribution, which we calculate numerically, into a prediction by summing probabilities for each of three outcomes: $r$ wins (states $L>0$), $r$ ties $b$, (state $L=0$), and $b$ wins (states $L>0$). In this way, we capture the information contained in magnitude of the current lead, which is lost when we simply predict that the current leader will win, regardless of lead size.

\begin{figure*}[t]
\centering
\includegraphics[scale=0.99]{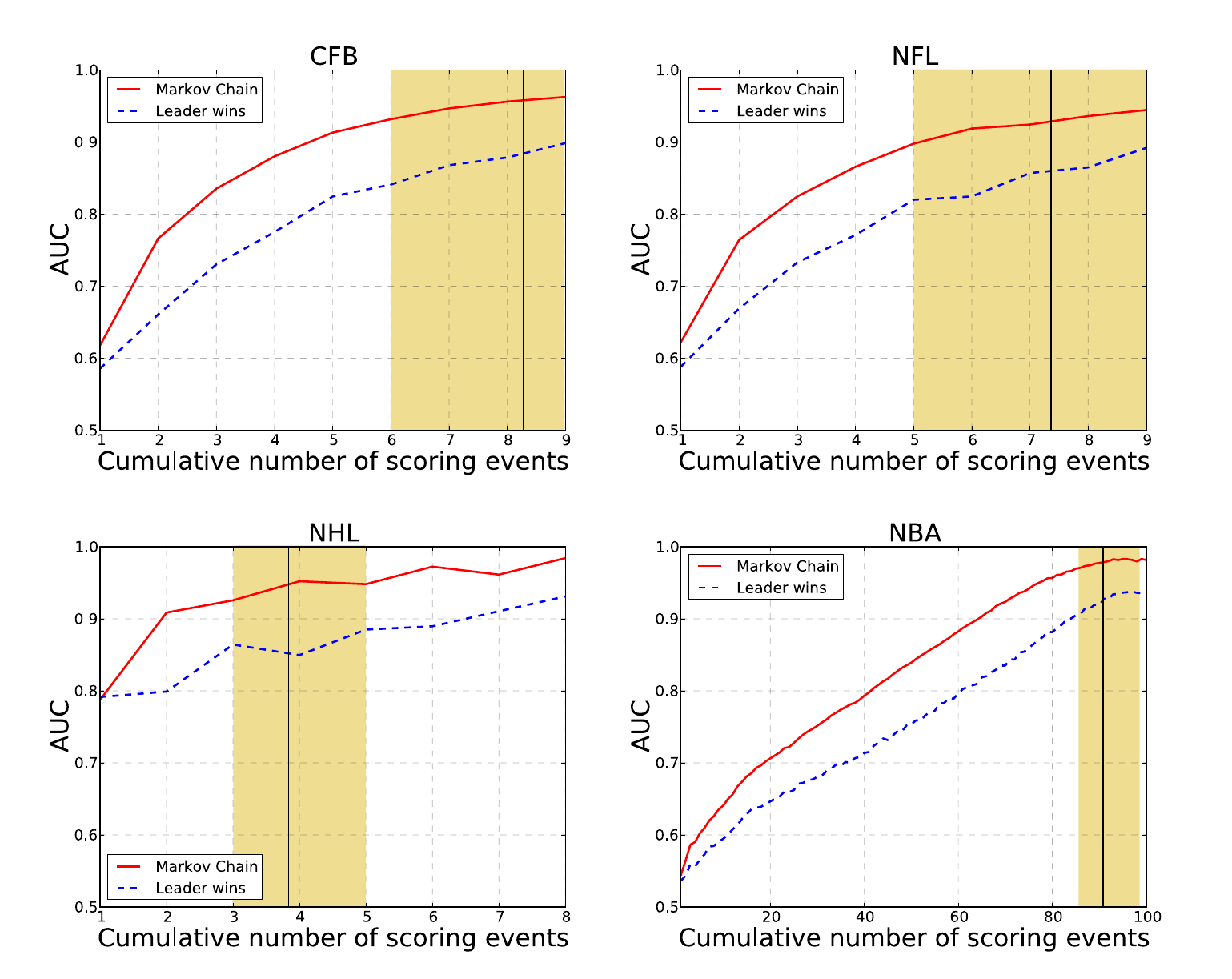}
\caption{\textbf{Predicting game outcome from dynamics.} Probability of correctly predicting the game winner (AUC) using the Markov chain model and the ``leader wins'' model for football, hockey, and basketball. The vertical line shows the average number of scoring events per game, and the highlighted region shows the middle 50\% of games around this value.}
\label{fig:pred}
\end{figure*}

We test the accuracy of the Markov chain using an out-of-sample prediction scheme, in which we repeatedly divide each sports' game data into a training set of a randomly selected 3/4 of all games and a test set of the remaining 1/4. From each training set, we estimate the empirical functions used in the model and compute the Markov chain's transition matrix. Then, across the games in each test set, we measure the mean fraction of times the Markov chain's prediction is correct. This fraction is equivalent to the popular AUC statistic~\cite{bradley:1997}, where AUC$\,=0.5$ denotes an accuracy no better than guessing.

Instead of evaluating the model at some arbitrarily selected time, we investigate how outcome predictability evolves over time. Specifically, we compute the AUC as a function of the cumulative number of scoring events in the game, using the empirically observed times and lead sizes in each test-set game to parameterize the model's predictions. When the number of cumulative events is small, game outcomes should be relatively unpredictable, and as the clock runs down, predictability should increase. To provide a reference point for the quality of these results, we also measure the AUC over time for a simple heuristic of predicting the winner as the team in the lead after the event.

Across all sports, we find that game outcome is highly predictable, even after only a small number of scoring events (Figure~\ref{fig:pred}). For instance, the winner of CFB and NFL games can be accurately chosen more than 60\% of the time after only a single scoring event, and this rate increases to more than 80\% by three events. NHL games are even more predictable, in part because they are very low-scoring games, and the winner may be accurately chosen roughly 80\% of the time after the first event. The fast rise of the AUC curve as a function of continued scoring in these sports likely reflects the role played by differences in latent team skill in producing large leads, which make outcomes more predictable (Figure~\ref{fig:scoring:lead}).
In contrast, NBA games are the least predictable, requiring more than 40 events before the AUC exceeds 80\%. This likely reflects the role of the ``restoring force'' (Figure~\ref{fig:scoring:lead}), which tends to make NBA games more unpredictable than we would expect from a simple model of scoring, and significantly more unpredictable than CFB, NFL or NHL games.

In all cases, the Markov chain substantially outperforms the ``leader wins'' heuristic, even in the low-scoring NHL games. This occurs in part because small leads are less informative than large leads for guessing the winner, and the heuristic does not distinguish between these.

\section{Discussion}
Although there is increasing interest in quantitative analysis and modeling in sports~\cite{arkes2011finally,bourbousson:seve:mcgarry:2012,de2011basketball,everson2008composite,neiman2011reinforcement}, many questions remain about what patterns or principles, if any, cut across different sports, what basic dynamical processes provide good models of within-game events, and the degree to which the outcomes of games may be predicted from within-game events alone. 
The comprehensive database of scoring events we use here to investigate such questions is unusual for both its scope (every league game over 9--10 seasons), its breadth (covering four sports), and its depth (timing and attribution information on every point in every game). As such, it offers a number of new opportunities to study competition in general, and sports in particular.

Across college (American) football (CFB), professional (American) football (NFL), professional hockey (NHL) and professional basketball (NBA) games, we find a number of common patterns in both the tempo and balance of scoring events. First, the timing of events in all four sports is remarkably well-approximated by a simple Poisson process (Figures~\ref{fig:timing:events} and~\ref{fig:timing:iat}), in which each second of gameplay produces a scoring event independently, with a probability that varies only modestly over the course of a game (Figure~\ref{fig:timing:rate}). These variations, however, follow a common three-phase pattern, in which a relatively constant rate is depressed at the beginning of a scoring period, and increases dramatically in the final few seconds of the period. The excellent agreement with a Poisson process implies that teams employ very few strategically-chosen chains of events or time-sensitive strategies in these games, except in a period's end-phase, when the incentive to score in elevated. These results provide further support to some past analyses~\cite{ayton2004hot,gabel:redner:2012}, while contrasting with others~\cite{yaari:david:2012,yaari:eisenman:2011}, showing no evidence for the popular notion of ``hot hands,'' in which scoring once increases the chance of scoring again soon.

Second, we find a common pattern of imbalanced scoring between teams in CFB, NFL and NHL games, relative to an ideal model in which teams are equally likely to win each scoring event (Figure~\ref{fig:scoring:balance}). CFB games are much less balanced than NFL games, suggesting that the transition from college to professional tends to reduce the team skill differences that generate lopsided scoring. This reduction in variance is likely related both to only the stronger college-level players successfully moving up into the professional teams, and in the way the NFL Draft tends to distributed the stronger new players to the weaker teams.

Furthermore, we find that all three of these sports exhibit a pattern in which lead sizes tend to increase over time. That is, the probability of scoring while in the lead tends to be larger the greater the lead size (Figure~\ref{fig:scoring:lead}), in contrast to the ideal model in which lead sizes increase or decrease with equal probability. As with overall scoring balance, the size of this effect in CFB games is much larger (about 2.5 times larger) than in NFL games, and is consistent with a reduction in the variance of the distribution of skill across teams. That is, NFL teams are generally closer in team skill than CFB teams, and this produces gameplay that is much less predictable. Both of these patterns are consistent with a kind of Bradley-Terry-type model in which each scoring event is a contest between the teams.

NBA games, however, present the opposite pattern: team scores are much closer than we would expect from the ideal model, and the probability of scoring while in the lead effectively \textit{decreases} as the lead size grows (Figure~\ref{fig:scoring:lead}; a pattern originally identified by~\cite{gabel:redner:2012}). This pattern produces a kind of ``restoring force'' such that leads tend to shrink until they turn into ties, producing games that are substantially more unpredictable. Unlike the pattern in CFB, NFL and NHL, no distribution of latent team skills, under a Bradley-Terry-type model, can produce this kind of negative correlation between the probability of scoring and lead size.

Recently, \cite{berger:2011} analyzed similar NBA game data and argued that increased psychological motivation drives teams that are slightly behind (e.g., by one point at halftime) to win the game more often than not. That is, losing slightly is good for winning. Our analysis places this claim in a broader, more nuanced context. The effective restoring force is superficially consistent with the belief that losing in NBA games is ``good'' for the team, as losing does indeed empirically increase the probability of scoring. However, we find no such effect in CFB, NFL or NHL games (Figure~\ref{fig:scoring:lead}), suggesting either that NBA players are more poorly motivated than players in other team sports or that some other mechanism explains the pattern. 

One such mechanism is for NBA teams to employ strategies associated with substituting weaker players for stronger ones when they hold various leads, e.g., to allow their best players to rest or avoid injury, manage floor spacing and offensive/defensive combinations, etc., and then reverse the process when the other team leads. In this way, a team will play more weakly when it leads, and more strongly when it is losing, because of personnel changes alone rather than changes in morale or effort. If teams have different thresholds for making such substitutions, and differently skilled best players, the averaging across these differences would produce the smooth pattern observed in the data. Such substitutions are indeed common in basketball games, while football and hockey teams are inherently less able to alter their effective team skill through such player management, which may explain the restoring force's presence in NBA games and its absence in CFB, NFL or NHL games. It would be interesting to determine whether college basketball games exhibit the same restoring force, and the personnel management hypothesis could be tested by estimating the on-court team's skill as a function of lead size.

The observed patterns we find in the probability of scoring while in the lead are surprisingly accurate at reproducing the observed variation in lead-size dynamics in these sports (Figure~\ref{fig:sim:lead}), and suggest that this one pattern provides a compact and mostly accurate summary of the within-game scoring dynamics of a sport. However, we do not believe these patterns indicate the presence of any feedbacks, e.g., ``momentum'' or cumulative advantage~\cite{price1976general}. Instead, for CFB, NFL and NHL games, this pattern represents the the distribution of latent team skills, while for NBA games, it represents strategic decisions about which players are on the court as a function of lead size. 

This pattern also makes remarkably good predictions about the overall outcome of games, even when given information about only the first $\ell$ scoring events. Under a controlled out-of-sample test, we found that CFB, NFL and NHL games are highly predictable, even after only a few events. In contrast, NBA games were significantly less predictable, although reasonable predictions here can still be made, despite the impact of the restoring force.

Given the popularity of betting on sports, it is an interesting question as to whether our model produces better or worse predictions than those of established odds-makers. To explore this question, we compared our model against two such systems, the online live-betting website Bovada%
\footnote{See {\tt https://live.bovada.lv}. Only data on NBA games were available.} 
and the odds-maker website Sports Book Review (SBR).%
\footnote{See {\tt http://www.sbrforum.com/betting-odds}}
Neither site provided comprehensive coverage or systematic access, and so our comparison was necessarily limited to a small sample of games. Among these, however, our predictions were very close to those of Bovada, and, after 20\% of each game's events had occurred, were roughly 10\% more accurate than SBR's money lines across all sports. Although the precise details are unknown for how these commercial odds were set, it seems likely that they rely on many details omitted by our model, such as player statistics, team histories, team strategies and strengths, etc. In contrast, our model uses only information relating to the basic scoring dynamics within a sport, and knows nothing about individual teams or game strategies. In that light, its accuracy is impressive.

These results suggest several interesting directions for future work. For instance, further elucidating the connection between team skill and the observed scoring patterns would provide an important connection between within-game dynamics and team-specific characteristics. These, in turn, could be estimated from player-level characteristics to provide a coherent understanding of how individuals cooperate to produce a team and how teams compete to produce dynamics. Another missing piece of the dynamics puzzle is the role played by the environment and the control of space for creating scoring opportunities. Recent work on online games with heterogeneous environments suggests that these spatial factors can have large impact on scoring tempo and balance~\cite{merritt2013environmental}, but time series data on player positions on the field would further improve our understanding. Finally, our data omit many aspects of gameplay, including referee calls, timeouts, fouls, etc., which may provide for interesting strategic choices by teams, e.g., near the end of the game, as with clock management in football games. Progress on these and other questions would shed more light on the fundamental question of how much of gameplay may be attributed to skill versus luck.

Finally, our results demonstrate that common patterns and processes do indeed cut across seemingly distinct sports, and these patterns provide remarkably accurate descriptions of the events within these games and predictions of their outcomes. However, many questions remain unanswered, particularly as to what specific mechanisms generate the modest deviations from the basic patterns that we observe in each sport, and how exactly teams exerting such great efforts against each other can conspire to produce gameplay so reminiscent of simple stochastic processes. We look forward to future work that further investigates these questions, which we hope will continue to leverage the powerful tools and models of dynamical systems, statistical physics, and machine learning with increasingly detailed data on competition.

\begin{acknowledgments}
We thank Dan Larremore, Christopher Aicher, Joel Warner, Mason Porter, Peter Mucha, Pete McGraw, Dave Feldman, Sid Redner, Alan Gabel, Dave Feldman, Owen Newkirk, Oskar Burger, Rajiv Maheswaran and Chris Meyer for helpful conversations. This work was supported in part by the James S. McDonnell Foundation.
\end{acknowledgments}


\begin{thebibliography}{43}
\ifx \bisbn   \undefined \def \bisbn  #1{ISBN #1}\fi
\ifx \binits  \undefined \def \binits#1{#1}\fi
\ifx \bauthor  \undefined \def \bauthor#1{#1}\fi
\ifx \batitle  \undefined \def \batitle#1{#1}\fi
\ifx \bjtitle  \undefined \def \bjtitle#1{#1}\fi
\ifx \bvolume  \undefined \def \bvolume#1{\textbf{#1}}\fi
\ifx \byear  \undefined \def \byear#1{#1}\fi
\ifx \bissue  \undefined \def \bissue#1{#1}\fi
\ifx \bfpage  \undefined \def \bfpage#1{#1}\fi
\ifx \blpage  \undefined \def \blpage #1{#1}\fi
\ifx \burl  \undefined \def \burl#1{\textsf{#1}}\fi
\ifx \doiurl  \undefined \def \doiurl#1{\textsf{#1}}\fi
\ifx \betal  \undefined \def \betal{\textit{et al.}}\fi
\ifx \binstitute  \undefined \def \binstitute#1{#1}\fi
\ifx \binstitutionaled  \undefined \def \binstitutionaled#1{#1}\fi
\ifx \bctitle  \undefined \def \bctitle#1{#1}\fi
\ifx \beditor  \undefined \def \beditor#1{#1}\fi
\ifx \bpublisher  \undefined \def \bpublisher#1{#1}\fi
\ifx \bbtitle  \undefined \def \bbtitle#1{#1}\fi
\ifx \bedition  \undefined \def \bedition#1{#1}\fi
\ifx \bseriesno  \undefined \def \bseriesno#1{#1}\fi
\ifx \blocation  \undefined \def \blocation#1{#1}\fi
\ifx \bsertitle  \undefined \def \bsertitle#1{#1}\fi
\ifx \bsnm \undefined \def \bsnm#1{#1}\fi
\ifx \bsuffix \undefined \def \bsuffix#1{#1}\fi
\ifx \bparticle \undefined \def \bparticle#1{#1}\fi
\ifx \barticle \undefined \def \barticle#1{#1}\fi
\ifx \bconfdate \undefined \def \bconfdate #1{#1}\fi
\ifx \botherref \undefined \def \botherref #1{#1}\fi
\ifx \url \undefined \def \url#1{\textsf{#1}}\fi
\ifx \bchapter \undefined \def \bchapter#1{#1}\fi
\ifx \bbook \undefined \def \bbook#1{#1}\fi
\ifx \bcomment \undefined \def \bcomment#1{#1}\fi
\ifx \oauthor \undefined \def \oauthor#1{#1}\fi
\ifx \citeauthoryear \undefined \def \citeauthoryear#1{#1}\fi
\ifx \endbibitem  \undefined \def \endbibitem {}\fi
\ifx \bconflocation  \undefined \def \bconflocation#1{#1}\fi
\ifx \arxivurl  \undefined \def \arxivurl#1{\textsf{#1}}\fi
\csname PreBibitemsHook\endcsname

\bibitem{klaassen:magnus:2001}
\begin{barticle}
\bauthor{\bsnm{Klaassen}, \binits{F.J.G.M.}},
\bauthor{\bsnm{Magnus}, \binits{J.R.}}:
\batitle{Are points in tennis independent and identically distributed? evidence
  from a dynamic binary panel data model}.
\bjtitle{Journal of the American Statistical Association}
\bvolume{96},
\bfpage{500}--\blpage{509}
(\byear{2001})
\end{barticle}
\endbibitem

\bibitem{albert2005anthology}
\begin{bbook}
\bauthor{\bsnm{Albert}, \binits{J.}},
\bauthor{\bsnm{Bennett}, \binits{J.}},
\bauthor{\bsnm{Cochran}, \binits{J.J.}}:
\bbtitle{Anthology of Statistics in Sports}
vol. \bseriesno{16}.
\bpublisher{Society for Industrial Mathematics},
\blocation{Philadelphia, PA}
(\byear{2005})
\end{bbook}
\endbibitem

\bibitem{ben2005most}
\begin{barticle}
\bauthor{\bsnm{Ben-Naim}, \binits{E.}},
\bauthor{\bsnm{Vazquez}, \binits{F.}},
\bauthor{\bsnm{Redner}, \binits{S.}}:
\batitle{What is the most competitive sport?}
\bjtitle{Journal of Korean Physics Society}
\bvolume{50},
\bfpage{124}--\blpage{126}
(\byear{2007})
\end{barticle}
\endbibitem

\bibitem{thomas2007inter}
\begin{botherref}
\oauthor{\bsnm{Thomas}, \binits{A.C.}}:
Inter-arrival times of goals in ice hockey.
Journal of Quantitative Analysis in Sports
\textbf{3}(3)
(2007)
\end{botherref}
\endbibitem

\bibitem{duch:etal:2010}
\begin{barticle}
\bauthor{\bsnm{Duch}, \binits{J.}},
\bauthor{\bsnm{Waitzman}, \binits{J.S.}},
\bauthor{\bsnm{Amaral}, \binits{L.A.N.}}:
\batitle{Quantifying the performance of individual players in a team activity}.
\bjtitle{PLOS ONE}
\bvolume{5},
\bfpage{10937}
(\byear{2010})
\end{barticle}
\endbibitem

\bibitem{heuer:etal:2010}
\begin{barticle}
\bauthor{\bsnm{Heuer}, \binits{A.}},
\bauthor{\bsnm{M\"uller}, \binits{C.}},
\bauthor{\bsnm{Rubner}, \binits{O.}}:
\batitle{Soccer: Is scoring goals a predictable {Poissonian} process?}
\bjtitle{Eur.\ Phys.\ Lett.}
\bvolume{89},
\bfpage{38007}
(\byear{2010})
\end{barticle}
\endbibitem

\bibitem{buttrey2011estimating}
\begin{barticle}
\bauthor{\bsnm{Buttrey}, \binits{S.E.}},
\bauthor{\bsnm{Washburn}, \binits{A.R.}},
\bauthor{\bsnm{Price}, \binits{W.L.}}:
\batitle{Estimating {NHL} scoring rates}.
\bjtitle{Journal of Quantitative Analysis in Sports}
\bvolume{7}(\bissue{3}),
\bfpage{24}
(\byear{2011})
\end{barticle}
\endbibitem

\bibitem{radicchi:2011}
\begin{barticle}
\bauthor{\bsnm{Radicchi}, \binits{F.}}:
\batitle{Who is the best player ever? {A} complex network analysis of the
  history of professional tennis}.
\bjtitle{PLOS ONE}
\bvolume{6},
\bfpage{17249}
(\byear{2011})
\end{barticle}
\endbibitem

\bibitem{radicchi:2012}
\begin{barticle}
\bauthor{\bsnm{Radicchi}, \binits{F.}}:
\batitle{Universality, limits and predictability of gold-medal performances at
  the {O}lympics games}.
\bjtitle{PLOS ONE}
\bvolume{7},
\bfpage{40335}
(\byear{2012})
\end{barticle}
\endbibitem

\bibitem{gabel:redner:2012}
\begin{botherref}
\oauthor{\bsnm{Gabel}, \binits{A.}},
\oauthor{\bsnm{Redner}, \binits{S.}}:
Random walk picture of basketball scoring.
Journal of Quantitative Analysis in Sports
\textbf{8}
(2012)
\end{botherref}
\endbibitem

\bibitem{goldman2012effort}
\begin{bchapter}
\bauthor{\bsnm{Goldman}, \binits{M.}},
\bauthor{\bsnm{Rao}, \binits{J.M.}}:
\bctitle{Effort vs. concentration: The asymmetric impact of pressure on {NBA}
  performance}.
In: \bbtitle{Proceedings MIT Sloan Sports Analytics Conference},
pp. \bfpage{1}--\blpage{10}
(\byear{2012})
\end{bchapter}
\endbibitem

\bibitem{yaari:david:2012}
\begin{barticle}
\bauthor{\bsnm{Yaari}, \binits{G.}},
\bauthor{\bsnm{David}, \binits{G.}}:
\batitle{``{Hot} hand'' on strike: Bowling data indicates correlation to recent
  past results, not causality}.
\bjtitle{PLOS ONE}
\bvolume{7},
\bfpage{30112}
(\byear{2012})
\end{barticle}
\endbibitem

\bibitem{myerson:1997}
\begin{bbook}
\bauthor{\bsnm{Myerson}, \binits{R.B.}}:
\bbtitle{Game Theory: Analysis of Conflict}.
\bpublisher{Harvard University Press},
\blocation{Cambridge MA}
(\byear{1997})
\end{bbook}
\endbibitem

\bibitem{palacios2003professionals}
\begin{barticle}
\bauthor{\bsnm{Palacios-Huerta}, \binits{I.}}:
\batitle{Professionals play minimax}.
\bjtitle{The Review of Economic Studies}
\bvolume{70}(\bissue{2}),
\bfpage{395}--\blpage{415}
(\byear{2003})
\end{barticle}
\endbibitem

\bibitem{walker2001minimax}
\begin{barticle}
\bauthor{\bsnm{Walker}, \binits{M.}},
\bauthor{\bsnm{Wooders}, \binits{J.}}:
\batitle{Minimax play at {Wimbledon}}.
\bjtitle{The American Economic Review}
\bvolume{91}(\bissue{5}),
\bfpage{1521}--\blpage{1538}
(\byear{2001})
\end{barticle}
\endbibitem

\bibitem{romer:2006}
\begin{barticle}
\bauthor{\bsnm{Romer}, \binits{D.}}:
\batitle{Do firms maximize? {Evidence} from professional football}.
\bjtitle{Journal of Political Economy}
\bvolume{114}(\bissue{2}),
\bfpage{340}--\blpage{365}
(\byear{2006})
\end{barticle}
\endbibitem

\bibitem{reed:hughes:2006}
\begin{barticle}
\bauthor{\bsnm{Reed}, \binits{D.}},
\bauthor{\bsnm{Hughes}, \binits{M.}}:
\batitle{An exploration of team sport as a dynamical system}.
\bjtitle{International Journal of Performance Analysis in Sport}
\bvolume{6}(\bissue{2}),
\bfpage{114}--\blpage{125}
(\byear{2006})
\end{barticle}
\endbibitem

\bibitem{galla:farmer:2013}
\begin{barticle}
\bauthor{\bsnm{Galla}, \binits{T.}},
\bauthor{\bsnm{Farmer}, \binits{J.D.}}:
\batitle{Complex dynamics in learning complicated games}.
\bjtitle{Proceedings of the National Academy of Sciences USA}
\bvolume{110},
\bfpage{1232}--\blpage{1236}
(\byear{2013})
\end{barticle}
\endbibitem

\bibitem{ayton2004hot}
\begin{barticle}
\bauthor{\bsnm{Ayton}, \binits{P.}},
\bauthor{\bsnm{Fischer}, \binits{I.}}:
\batitle{The hot hand fallacy and the gambler's fallacy: Two faces of
  subjective randomness?}
\bjtitle{Memory \& Cognition}
\bvolume{32}(\bissue{8}),
\bfpage{1369}--\blpage{1378}
(\byear{2004})
\end{barticle}
\endbibitem

\bibitem{balkundi:harrison:2006}
\begin{barticle}
\bauthor{\bsnm{Balkundi}, \binits{P.}},
\bauthor{\bsnm{Harrison}, \binits{D.A.}}:
\batitle{Ties, leaders, and time in teams: Strong inference about network
  structure's effects on team viability and performance}.
\bjtitle{Academy of Management Journal}
\bvolume{49},
\bfpage{49}--\blpage{68}
(\byear{2006})
\end{barticle}
\endbibitem

\bibitem{berger:2011}
\begin{barticle}
\bauthor{\bsnm{Berger}, \binits{J.}},
\bauthor{\bsnm{Pope}, \binits{D.}}:
\batitle{Can losing lead to winning?}
\bjtitle{Management Science}
\bvolume{57}(\bissue{5}),
\bfpage{817}--\blpage{827}
(\byear{2011})
\end{barticle}
\endbibitem

\bibitem{vergin2000winning}
\begin{botherref}
\oauthor{\bsnm{Vergin}, \binits{R.C.}}:
Winning streaks in sports and the misperception of momentum.
Journal of Sport Behavior
\textbf{23}
(2000)
\end{botherref}
\endbibitem

\bibitem{merritt2013environmental}
\begin{barticle}
\bauthor{\bsnm{Merritt}, \binits{S.}},
\bauthor{\bsnm{Clauset}, \binits{A.}}:
\batitle{Environmental structure and competitive scoring advantages in team
  competitions}.
\bjtitle{Scientific Reports}
\bvolume{3},
\bfpage{3067}
(\byear{2013})
\end{barticle}
\endbibitem

\bibitem{barney:1991}
\begin{barticle}
\bauthor{\bsnm{Barney}, \binits{J.}}:
\batitle{Firm resources and sustained competitive advantage}.
\bjtitle{Journal of Management}
\bvolume{17},
\bfpage{99}--\blpage{120}
(\byear{1991})
\end{barticle}
\endbibitem

\bibitem{Note1}
\begin{botherref}
Data provided by STATS LLC, copyright 2014.
\end{botherref}
\endbibitem

\bibitem{yaari:eisenman:2011}
\begin{barticle}
\bauthor{\bsnm{Yaari}, \binits{G.}},
\bauthor{\bsnm{David}, \binits{G.}}:
\batitle{The hot (invisible?) hand: Can time sequence patterns of
  success/failure in sports be modeled as repeated independent trials}.
\bjtitle{PLOS ONE}
\bvolume{6},
\bfpage{24532}
(\byear{2011})
\end{barticle}
\endbibitem

\bibitem{boas:2006}
\begin{bbook}
\bauthor{\bsnm{Boas}, \binits{M.L.}}:
\bbtitle{Mathematical Methods in the Physical Sciences},
\bedition{3}rd edn.
\bpublisher{John Wiley \& Sons, Inc.},
\blocation{Hoboken, NJ}
(\byear{2006})
\end{bbook}
\endbibitem

\bibitem{box2011time}
\begin{bbook}
\bauthor{\bsnm{Box}, \binits{G.E.P.}},
\bauthor{\bsnm{Jenkins}, \binits{G.M.}},
\bauthor{\bsnm{Reinsel}, \binits{G.C.}}:
\bbtitle{Time Series Analysis: Forecasting and Control}.
\bpublisher{John Wiley \& Sons, Inc.},
\blocation{Hoboken, NJ}
(\byear{2013})
\end{bbook}
\endbibitem

\bibitem{thompson:2010}
\begin{bchapter}
\bauthor{\bsnm{Thompson}, \binits{P.}}:
\bctitle{Learning by doing}.
In: \beditor{\bsnm{Hall}, \binits{B.}},
\beditor{\bsnm{Rosenberg}, \binits{N.}} (eds.)
\bbtitle{Handbook of Economics of Technical Change},
pp. \bfpage{429}--\blpage{476}.
\bpublisher{Elsevier/North-Holland},
\blocation{Philadelphia PA}
(\byear{2010})
\end{bchapter}
\endbibitem

\bibitem{bradley1952rank}
\begin{barticle}
\bauthor{\bsnm{Bradley}, \binits{R.A.}},
\bauthor{\bsnm{Terry}, \binits{M.E.}}:
\batitle{Rank analysis of incomplete block designs: I. {The} method of paired
  comparisons}.
\bjtitle{Biometrika}
\bvolume{39}(\bissue{3/4}),
\bfpage{324}--\blpage{345}
(\byear{1952})
\end{barticle}
\endbibitem

\bibitem{bradley:1997}
\begin{barticle}
\bauthor{\bsnm{Bradley}, \binits{A.P.}}:
\batitle{The use of the area under the {ROC} curve in the evaluation of machine
  learning algorithms}.
\bjtitle{Pattern Recognition}
\bvolume{30}(\bissue{7}),
\bfpage{1145}--\blpage{1159}
(\byear{1997})
\end{barticle}
\endbibitem

\bibitem{arkes2011finally}
\begin{botherref}
\oauthor{\bsnm{Arkes}, \binits{J.}},
\oauthor{\bsnm{Martinez}, \binits{J.}}:
Finally, evidence for a momentum effect in the {NBA}.
Journal of Quantitative Analysis in Sports
\textbf{7}
(2011)
\end{botherref}
\endbibitem

\bibitem{bourbousson:seve:mcgarry:2012}
\begin{barticle}
\bauthor{\bsnm{Bourbousson}, \binits{J.}},
\bauthor{\bsnm{S\`eve}, \binits{C.}},
\bauthor{\bsnm{{McGarry}}, \binits{T.}}:
\batitle{Space-time coordination dynamics in basketball: Part 2. {The}
  interaction between the two teams}.
\bjtitle{Journal of Sports Sciences}
\bvolume{28}(\bissue{3}),
\bfpage{349}--\blpage{358}
(\byear{2012})
\end{barticle}
\endbibitem

\bibitem{de2011basketball}
\begin{barticle}
\bauthor{\bparticle{de} \bsnm{Sa{\'a}~Guerra}, \binits{Y.}},
\bauthor{\bsnm{Mart{\'\i}n~Gonz{\'a}lez}, \binits{J.M.}},
\bauthor{\bsnm{Sarmiento~Montesdeoca}, \binits{S.}},
\bauthor{\bsnm{Rodr{\'\i}guez~Ruiz}, \binits{D.}},
\bauthor{\bsnm{Arjonilla~L{\'o}pez}, \binits{N.}},
\bauthor{\bsnm{Garc{\'\i}a~Manso}, \binits{J.M.}}:
\batitle{Basketball scoring in {NBA} games: an example of complexity}.
\bjtitle{Journal of Systems Science and Complexity}
\bvolume{26}(\bissue{1}),
\bfpage{94}--\blpage{103}
(\byear{2013})
\end{barticle}
\endbibitem

\bibitem{everson2008composite}
\begin{barticle}
\bauthor{\bsnm{Everson}, \binits{P.}},
\bauthor{\bsnm{Goldsmith-Pinkham}, \binits{P.S.}}:
\batitle{Composite {Poisson} models for goal scoring}.
\bjtitle{Journal of Quantitative Analysis in Sports}
\bvolume{4}(\bissue{2}),
\bfpage{13}
(\byear{2008})
\end{barticle}
\endbibitem

\bibitem{neiman2011reinforcement}
\begin{barticle}
\bauthor{\bsnm{Neiman}, \binits{T.}},
\bauthor{\bsnm{Loewenstein}, \binits{Y.}}:
\batitle{Reinforcement learning in professional basketball players}.
\bjtitle{Nature Communications}
\bvolume{2},
\bfpage{569}
(\byear{2011})
\end{barticle}
\endbibitem

\bibitem{price1976general}
\begin{barticle}
\bauthor{\bsnm{Price}, \binits{D.d.S.}}:
\batitle{A general theory of bibliometric and other cumulative advantage
  processes}.
\bjtitle{Journal of the American Society for Information Science}
\bvolume{27}(\bissue{5}),
\bfpage{292}--\blpage{306}
(\byear{1976})
\end{barticle}
\endbibitem

\bibitem{Note2}
\begin{botherref}
See {\protect \tt https://live.bovada.lv}. Only data on NBA games were
  available.
\end{botherref}
\endbibitem

\bibitem{Note3}
\begin{botherref}
See {\protect \tt http://www.sbrforum.com/betting-odds}
\end{botherref}
\endbibitem

\bibitem{nflRules}
\begin{botherref}
\oauthor{\bsnm{{National Football Association}}}:
Official rules of the {National Football Association}
(2013).
accessed 1 July 2013.
Accessed 07-01-2013
\end{botherref}
\endbibitem

\bibitem{ncaaRules}
\begin{botherref}
\oauthor{\bsnm{{National Collegiate Athletic Association}}}:
Football, rules and interpretations
(2013).
accessed 1 July 2013.
Accessed 07-01-2013
\end{botherref}
\endbibitem

\bibitem{nhlRules}
\begin{botherref}
\oauthor{\bsnm{{National Hockey League}}}:
Official rules of the {National Hockey League}
(2013).
accessed 1 July 2013.
Accessed 07-01-2013
\end{botherref}
\endbibitem

\bibitem{nbaRules}
\begin{botherref}
\oauthor{\bsnm{{National Basketball Association}}}:
Official rules of the {National Basketball Association}
(2013).
accessed 1 July 2013.
Accessed 07-01-2013
\end{botherref}
\endbibitem

\end{thebibliography}


\begin{appendix}

\section{Game mechanics}
\label{appendix:rules}
Here we provide brief summaries of the game mechanics for each of the sports represented in our data set.
\\

\noindent \textbf{Professional and college football}\\
Both professional and college football games last 60 minutes, divided into four equal length ``quarters.'' Each of the two teams field 11 players, the identities of which are usually changed depending on whether a team has possession of the ball (offense) or not (defense). The field is a flat grass or turf surface 360 feet long and 160 feet wide. On either end of the field are two ``end zones,'' measuring 30 feet in length, one for each team. 

The offense is given a series of four attempts (``downs'') to move the football 10 yards downfield from the last valid ball position, each of which must occur within 40 seconds of the last attempt's end. Failure to move the ball the required distance results in the other team gaining possession. Points are scored by the team in possession when it moves the ball into the defensive team's ``end zone,'' (a touchdown, 6 points) or passes the ball through the defensive team's field goal (a field goal, 3 points). Scoring a touchdown provides the scoring team the opportunity for additional points, either through what would normally be a touchdown (2 points) or a field goal (1 point). Each team has three timeouts to use during gameplay, which are often used strategically near the end of the game. When time runs out, the team with the most points is declared the winner. For a complete description of professional and college level rules, see~\cite{nflRules} and~\cite{ncaaRules} respectively.
\\

\noindent \textbf{Professional hockey}\\
A professional hockey game lasts 60 minutes, divided into three equal length ``periods." A game is played between two teams, each composed of six players (five skaters and one goalkeeper), whose identities change periodically throughout the game. Teams compete on an ice rink, 200 feet long and 85 feet wide. On either end of the rink are two nets, 6 feet wide and 4 feet high, one for each team.

Players on the team controlling the puck work together to move it into the opposing team's net through a combination of strategic passes and shots. If the team is successful, a goal is scored and the team is awarded 1 point. The game plays continuously except after stoppages, which occur at minutes 6, 10, and 14, penalties, or goals. Each team has a single 30 second timeout that can be used at any point in the game. Teams use their timeouts to substitute players, adjust strategy and to provide the team with brief moments of rest during crucial periods of play. When time runs out, the team with the most points is deemed the winner. For a complete description of rules, see~\cite{nhlRules}.
\\

\noindent \textbf{Professional basketball}\\
A professional basketball game lasts 48 minutes, divided into four equal length ``quarters." Each of the two teams field five players, whose identities change throughout the game. The court is a flat wooden surface, 94 feet long and 50 feet wide. On either side of the court are two circular rims, known as baskets, measuring 18 inches in diameter, positioned 10 feet above the court surface, one for each team.

A team in possession of the basketball has a total of 24 seconds to make a shot that it either hits the opposing team's rim or goes through it. If time expires before the team attempts a shot, the opposing team gains possession of the basketball. Depending on game state and a player's court location, a successful shot (one that goes through the opposing team's rim) can be worth 1, 2, or 3 points. After scoring, the scoring team relinquishes possession of the basketball to the opposing team. Game play continues according to this procedure, except when the ball goes out of bounds or a foul is committed. Each team is awarded a single 20 second timeout per game half. Each team is also entitled to 6 more timeouts that may be used at anytime throughout the game, with the following restrictions: no more than 3 timeouts may be used during the final quarter and no more than 2 timeouts may be used within the final 2 minutes of play. These timeouts are used strategically to substitute players, control the speed of play, and facilitate the coordination and planning of complex plays. When time expires, the team that has accumulated the most points is deemed the winner. For a full description of game rules, see~\cite{nbaRules}.

\section{Points per scoring event}
\label{appendix:points:per:event}
Table~\ref{tab:pts} shows the distribution of points per scoring event, for each sport. Events in the NHL only generate a single point. Although events in the NBA generate 1, 2 or 3 points, the large majority of events (74\%) are worth 2 points, with the remaining events divided between 1- and 3-point shots.

Similarly, scoring events in both CFB and NFL games generally produce 7 points (touchdown with  extra point). Games in CFB games from those in NFL in producing many more field goals (3 points) and many fewer touchdowns with no extra point (6 points), which are the next most common events in both. The remaining point values are relatively uncommon: 8 points for touchdowns plus a 2 point conversion play, and 2 points for a safety, which occurs in three scenarios: (i) when a ball carrier is tackled in his team's own end zone; (ii) when the ball is deemed dead by referees in the end zone, or (iii) when the offensive team commits a foul play in its own end zone. Two point conversions occur when the scoring team elects to successfully pass or run the ball into the end zone instead of kicking the ball through the field goal after a touchdown.

\begin{table}[t!]
\begin{center}
\begin{tabular}{c|cccc}
point value & NFL & CFB & NHL & NBA \\ \hline
1 & - & - &  \bf{1.0000} & 0.0941 \\
2 & 0.0083 & 0.0113 & - & \bf{0.7373} \\
3 & 0.3055 & 0.1702 & - & 0.1647 \\
4 & - & - & - & 0.0029 \\
5 & - & - & - & 0.0009 \\
6 & 0.0308 & 0.0708 & - & 0.0001 \\
7 & \bf{0.6222} & \bf{0.7058} & - & - \\
8 & 0.0332 & 0.0419 & - & - \\ \hline
any & 1.0000 & 1.0000 & 1.0000 & 1.0000
\end{tabular}
\end{center}
\caption{Empirical distribution of all regulation scoring events over point values, by sport, with the modal value highlighted.}
\label{tab:pts}
\end{table}

\begin{figure*}[t!]
\centering
\begin{tabular}{cccc}
\includegraphics[scale=0.36]{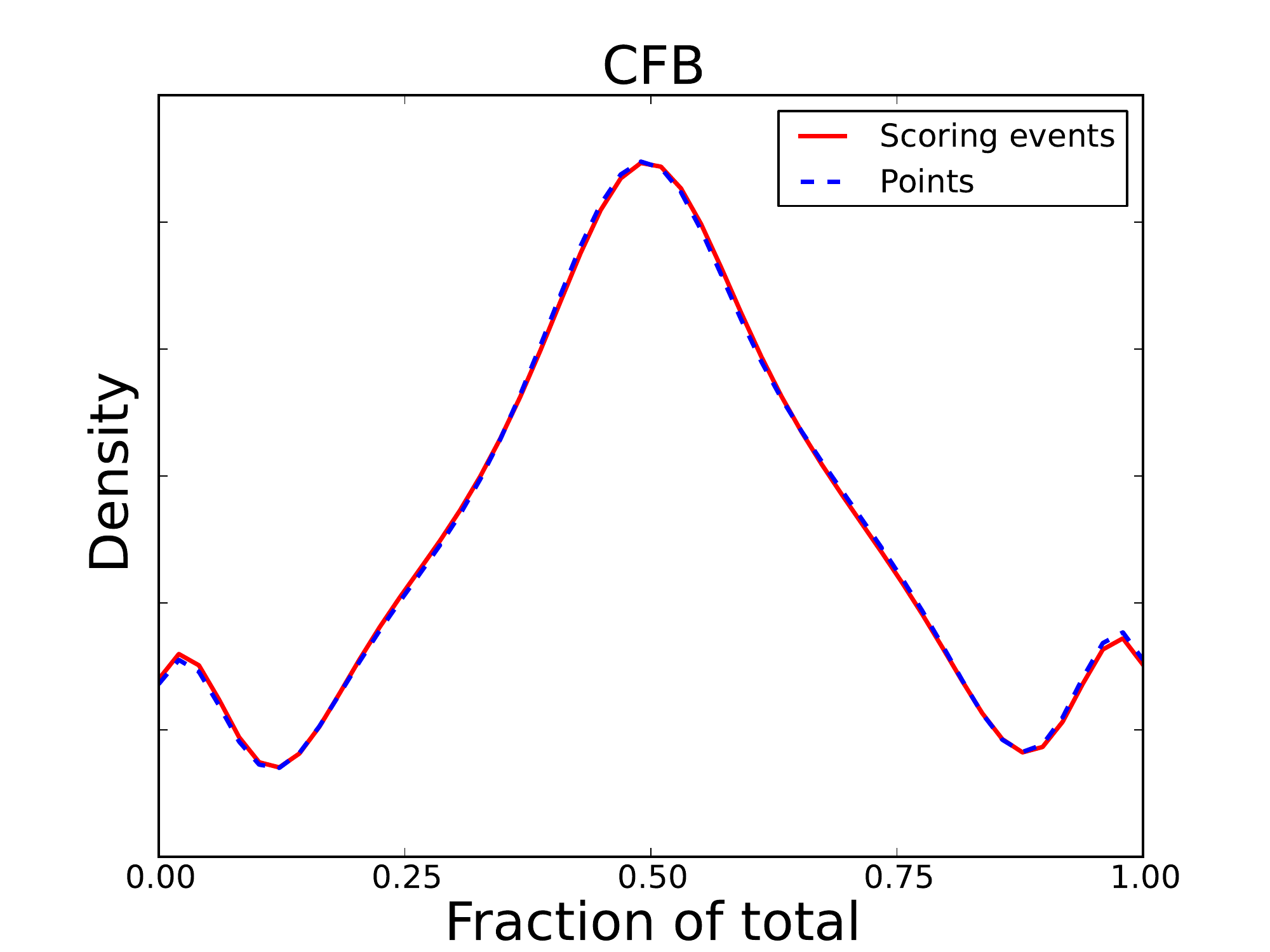} &
\includegraphics[scale=0.36]{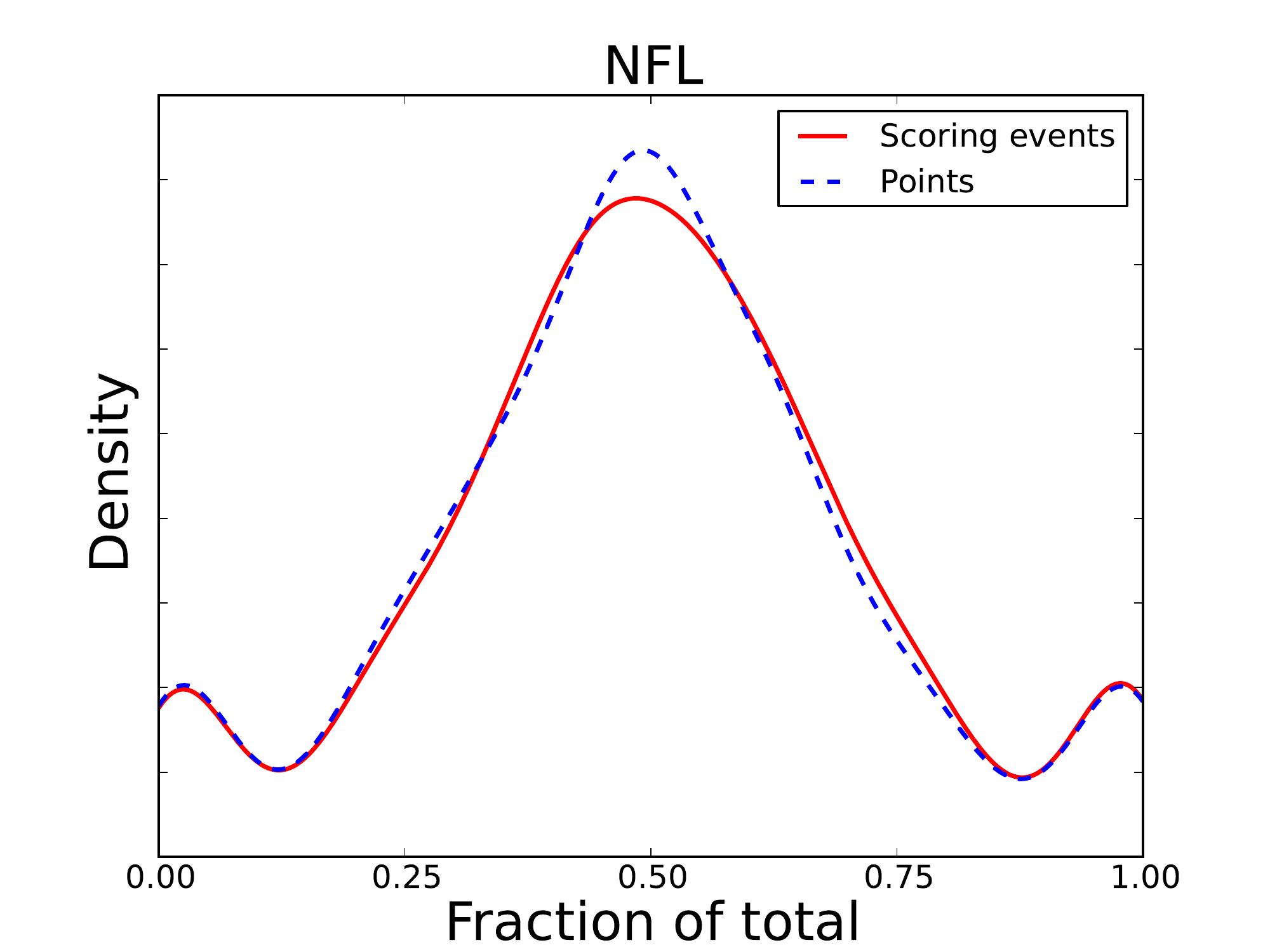} \\
\includegraphics[scale=0.36]{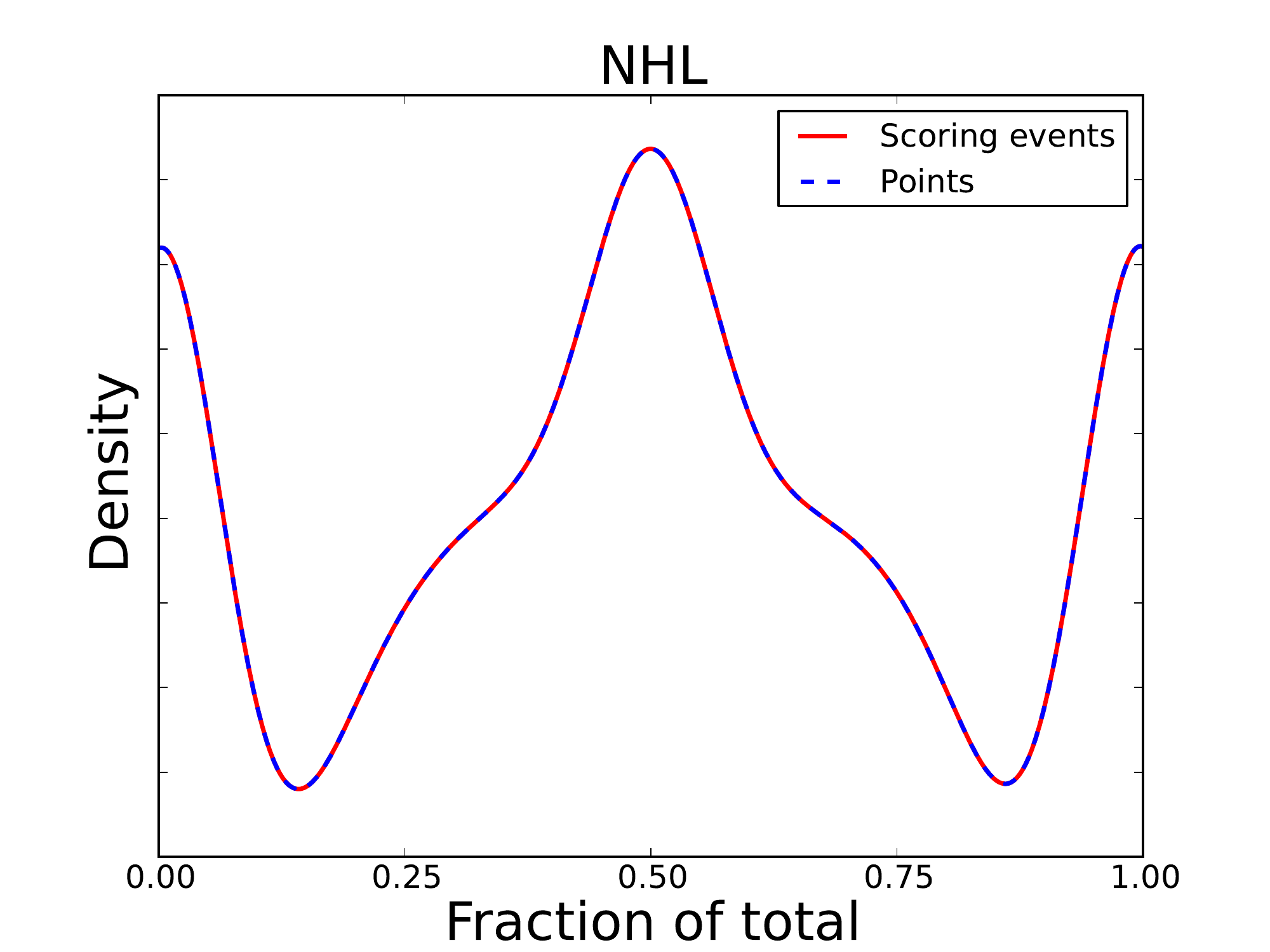} &
\includegraphics[scale=0.36]{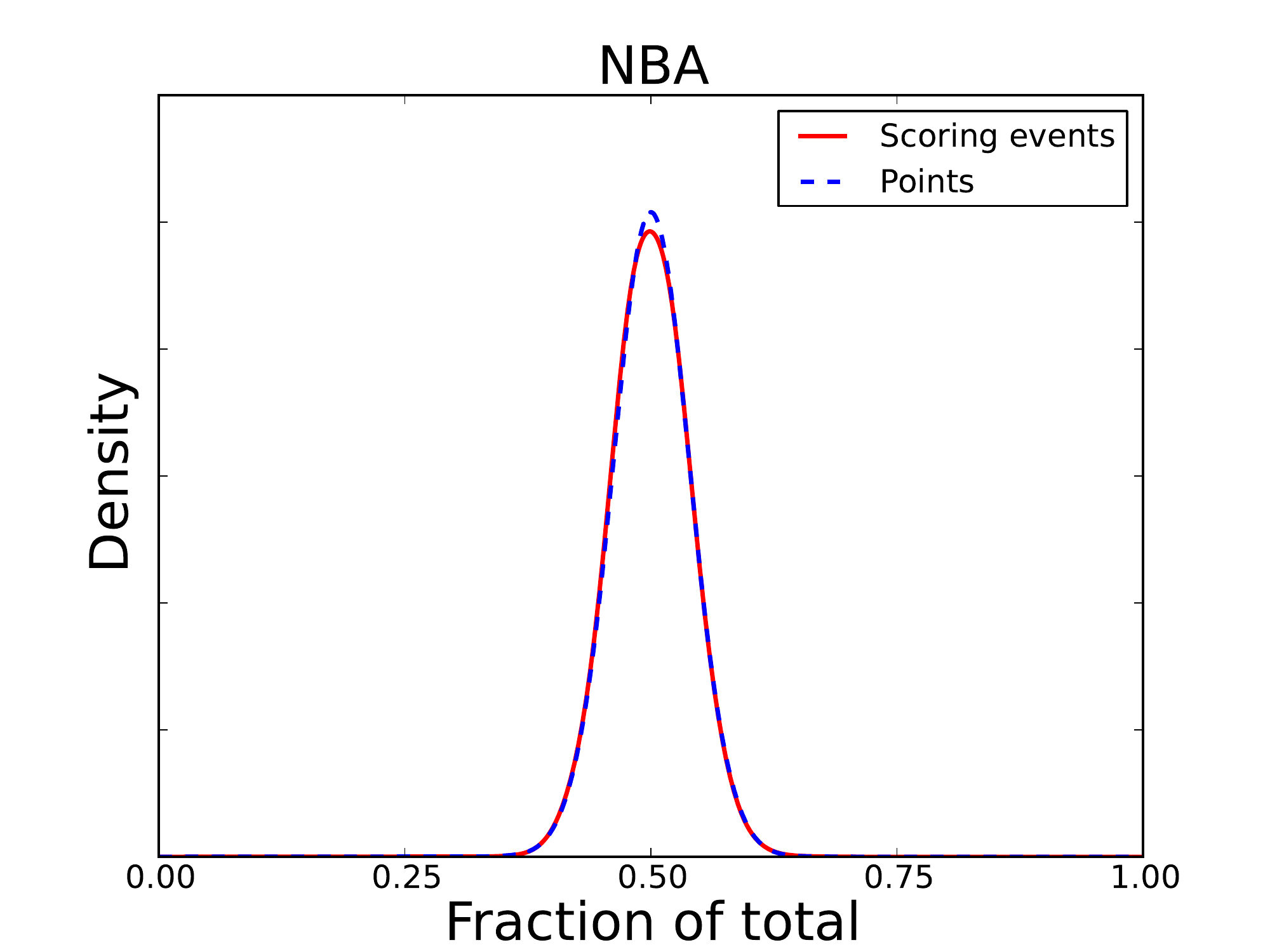}
\end{tabular}
\caption{Smoothed distributions for the empirical fraction of total points won by a team (solid line), for each sport, plus the empirical fraction of total scoring events (dashed line; from Figure~\ref{fig:scoring:balance}). The very close agreement indicates that only very rarely does the point-value of scoring events---instead of simply their number---determine the outcome of a game. }
\label{fig:scoring:pts}
\end{figure*}

Figure~\ref{fig:scoring:pts} shows the fraction of total points in each game that are won by a team, which agrees very closely with the fraction of total scoring events, from Figure~4 (main text). This agreement indicates that only very rarely does the value of the points associated with events ultimately determine the outcome of a game. Instead, the chief determinant is simply number of events. In NHL games, this must be true as every event is worth the same number of points. A slightly deviation around $1/2$ for NFL games, but not CFB games, indicates that very occasionally point values do matter.

\newpage

\end{appendix}

\begin{table*}[h!]
\begin{center}
\begin{small}
\begin{tabular}{p{7.9cm}|p{7.9cm}}
\textit{Question} & \textit{Answer} \\
\hline
Does scoring in games of different team sports follow common patterns? & Yes. The pattern of when points are scored and who gets them are remarkably similar across sports.\\ 
\hline
What is the common pattern? & Events occur randomly (a Poisson process). Which team wins the points is coin flip (a Bernoulli process) that depends on the relative skill difference of the teams on the field. \\
\hline
What might cause this pattern? & A strong focus on short-term maximization of scoring opportunities, while blocking the other team from the same. There is no evidence of strategic planning across plays, as in games like chess or Go. Teams largely react to events as they occur.\\
\hline
What determines how often scoring occurs? & Each sport has a characteristic rate (see Table~\ref{table:tempo:balance}), which increases dramatically at the end of scoring periods.\\
\hline
What determines who wins an event? & Skill and luck, in that order. \\
\hline
Do events early in a game influence events later in a game?    & No. Each scoring event or ``play'' is effectively independent, once we control for relative team skill (and lead size in basketball). Gameplay is effectively ``memoryless.''  \\
\hline
Can a team be ``hot,'' where they score in streaks?  & No. Just like players~\cite{ayton2004hot}, teams do not get ``hot.'' Scoring streaks are caused by getting lucky.\\
\hline
When is it easier or harder to score? & Every moment is equally easy or difficult. But, teams try harder at the end of a period. \\
\hline
Which sport is the most unpredictable? & Pro basketball, where lead sizes (spreads) tend to shrink back to zero. This tendency generates many ``ties'' as a game unfolds. \\
\hline
Do other sports exhibit this pattern? & No. Pro basketball is the only sport where the spread tends to shrink. In football and hockey, the spread tends to grow over time. \\
\hline
Does being behind help you win, as argued by~\cite{berger:2011}? & No. Being behind helps you lose. Being ahead and being lucky helps you win.
\end{tabular}
\end{small}
\end{center}
\caption{A summary of our results, in question-and-answer format.}
\label{table:q:and:a}
\end{table*}

\end{document}